\documentclass[11pt,a4paper]{article}
\usepackage{jcappub}
\usepackage[utf8]{inputenc}
\usepackage{amsmath,amssymb}
\usepackage{graphicx}
\usepackage{hyperref}
\usepackage{natbib}

\newcommand{\bq}{\textbf{q}}
\newcommand{\bx}{\textbf{x}}
\newcommand{\bPsi}{\boldsymbol{\Psi}}

\newcommand{\eq}[1]{Eq.~(\ref{#1})}

\newcommand{\fig}[1]{Figure~\ref{#1}}

\newcommand{\vb}[1]{\mathbf{#1}}
\def\ie{{\em i.e.}~}

\newcommand{\avg}[1]{\ensuremath{\left\langle \,#1\, \right\rangle}}

\def\cGpc{\, h^{-3} \, {\rm Gpc}^3}
\def\kMpc{\, h \, {\rm Mpc}^{-1}}
\def\icMpc{\, h^3 \, {\rm Mpc}^{-3}}

\def\dq{\mathrm{d}^3 q\,}
\def\ZA{e^{i \mathbf{k}\cdot \mathbf{q}}\,e^{-\frac{1}{2} k_i k_j A^{mm}_{ij}}}

\author[a]{Shi-Fan Chen}
\author[a,b]{Emanuele Castorina}
\author[a,b,c]{Martin White}
\affiliation[a]{Department of Physics, University of California,
Berkeley, CA 94720}
\affiliation[b]{Berkeley Center for Cosmological Physics, Berkeley, CA 94720}
\affiliation[c]{Department of Astronomy, University of California,
Berkeley, CA 94720}
\emailAdd{shifan\_chen@berkeley.edu}
\emailAdd{ecastorina@berkeley.edu}
\emailAdd{mwhite@berkeley.edu}

\title{Biased Tracers of Two Fluids in the Lagrangian Picture}

\abstract{We explore Lagrangian perturbation theory (LPT) for biased tracers in the presence of two fluids, focusing on the case of cold dark matter (CDM) and baryons. The presence of two fluids induces corrections to the Lagrangian bias expansion and tracer advection, both of which we formulate as expansions in the three linear modes of the Lagrangian equations of motion. We compute the linear-order two-fluid corrections in the Zeldovich approximation, finding that modifications to the bias expansion and tracer advection both enter as percent-level corrections over a large range of wavenumbers at low redshift and draw parallels with the Eulerian formalism. We then discuss nonlinear corrections in the two-fluid picture, and calculate contributions from the relative velocity effect ($\propto \vb{v}_r^2$) at one loop order. Finally, we conduct an exploratory Fisher analysis to assess the impact of two-fluid corrections on baryon acoustic oscillations (BAO) measurements, finding that while modest values of the relative bias parameters can introduce systematic biases in the measured BAO scale of up to $0.5\, \sigma$, fitting for these effects as additional parameters increases the error bar by less than $30\%$ across a wide range of bias values.} 

\begin{document}

\maketitle

\section{Introduction}
Observations of the large-scale structure (LSS) of the universe allow us to shed light on areas of physics ranging from galaxy formation and evolution to fundamental physics. A prime target of present and future LSS surveys is the measurement of baryon-acoustic oscillations (BAO) -- the imprints of sound waves in the baryon-photon fluid observed in the cosmic microwave background (CMB) on the observed clustering of galaxies -- which can be used as a standard ruler to constrain the expansion of the universe \cite{Weinberg13}. Upcoming surveys such as DESI \cite{DESI}, EUCLID \cite{EUCLID18} and WFIRST \cite{WFIRST18} will provide BAO measurements with higher-than-ever precision, and even more futuristic BAO surveys such as a Stage II 21-cm experiment \cite{CVDE-21cm} have been proposed. These next-generation observational campaigns will require us to model the LSS with unprecedented accuracy, at the sub-\% level.

One area of recent interest in the field of LSS has been in accounting for the effects induced by the existence of multiple species (cold dark matter, baryons, neutrinos), with similar but distinct clustering properties, using analytic methods. Studies of the perturbative approach to structure formation have traditionally grouped all nonrelativistic species into a ``total matter'' fluid, whose gravitational collapse is the dominant source of structure on cosmological scales in the late-time universe, but many authors have recently extended these techniques to include neutrinos \cite{Saito08,Shoji10,Blas14,Castorina15,Senatore17} and baryons \cite{Dalal10,Somogyi10, Tseliakhovich10,Bernardeau13,Lewandowski15,Schmidt16,Schmidt17} in the Eulerian framework of Standard Perturbation Theory (SPT). In parallel, the response of galaxy and halo formation to the existence of multiple fluid species has also been subject of extensive investigation \cite{Dalal10,Yoo11,Castorina14,LoVerde14,Munoz,Blazek16,Schmidt16,Schmidt17}. Of particular interest are the present-day imprint of relative perturbations between baryons and dark matter on large scales which, being seeded in the same epoch and at the same scales as the baryon acoustic oscillations, has the potential to confound future BAO measurements \cite{Dalal10,Blazek16,Beutler2017,Slepian}. While these relative perturbations do not grow significantly in time (and relative velocities in fact decay) and are thus small compared to the total-matter growing mode at late times, they amount to coherent supersonic flows post-recombination and could have significant effects on the formation of the first halos and galaxies \cite{Tseliakhovich10,Dalal10}, which are the progenitors of the objects we observe today.

The goal of this work is to formulate perturbation theory and galaxy bias in the presence of multiple fluids within the Lagrangian framework, with a particular focus on the two-fluid baryon-dark matter scenario. Our work is a direct extension of the aforementioned SPT calculations. While Lagrangian Peturbation Theory (LPT) is order-by-order equivalent to SPT, it seamlessly allows a consistent treatment of large scales bulk flows, which are responsible for the final shape and position of the BAO features in the correlation functions or power spectrum \cite{PadWhi09,Sherwin12,McCSza12,Whi14,PorSenZal14,Vlah16,TrevisanSen}.  The theory can also be extended to handle density field `reconstruction' \cite{ESSS07,PWC09,Noh09,Whi15,Sherwin19}.  These features make LPT a natural language for investigating possible distortions to the BAO feature.

This paper is organized as follows. In Section~\ref{sec:linear_eqs}, we introduce the linear Lagrangian equations of motion and discuss the role of non-gravitational forces such as Compton drag with the CMB. Modifications to Lagrangian galaxy bias and advection in the two-fluid limit are then introduced in Section~\ref{sec:bias}. In Section~\ref{sec:zeldovich}, we employ the results of the preceding two sections and calculate the lowest-order two-fluid corrections to the galaxy power spectrum in the Zeldovich approximation. Cross spectra and subtleties in the IR resummation are briefly discussed in Section~\ref{sec:cross_spectra}. In Section~\ref{sec:fisher} we take up whether the calculated two-fluid corrections can significantly bias BAO measurements, arguing that any such biases can be mitigated by simultaneously fitting for these easily-characterizeable effects. Our conclusions are summarized in Section~\ref{sec:conclusions}.

\section{Linear Equations of Motion in Lagrangian Space}
\label{sec:linear_eqs}

In the Lagrangian picture, fluid dynamics is encoded in the displacements $\Psi_\sigma(q)$ of fluid elements of each species, $\sigma$, originally situated at Lagrangian positions $\textbf{q}$, such that their Eulerian positions at conformal time $\tau$ ($d\tau = a^{-1} dt$) are given by \cite{Zel70,Whi14,Ber02}
\begin{equation}
   \textbf{x}_\sigma(\textbf{q},\tau) = \textbf{q} + \bPsi_\sigma(\textbf{q},\tau).
\end{equation}
The subscript $\sigma = \{c, b\}$ denotes the species, either cold dark matter (CDM) or baryons, respectively, whose motion we are tracking. Assuming that initial displacements are infinitesimally small compared to those at the redshifts of interest, the overdensity, $\delta_\sigma$, of each species at Eulerian position $x$ can be solved for via mass conservation
\begin{equation}
    1 + \delta_\sigma(\textbf{x},\tau) = \int d^3q \; \delta_D(\textbf{x} - \textbf{q} - \bPsi_\sigma(\textbf{q},\tau)) = \int d^3q \, \frac{d^3k}{(2\pi)^3} \; e^{i\textbf{k} \cdot (\textbf{x} - \textbf{q} - \bPsi_\sigma(\textbf{q},\tau))},
    \label{eqn:dens_eq}
\end{equation}
where $\delta_D$ is the Dirac delta function. Taylor expanding to first order in displacements yields the familiar result that $\delta_\sigma(x) = - \nabla \cdot \bPsi_\sigma(q)$, but, as seen in Equation~\ref{eqn:dens_eq}, one feature of working in the Lagrangian picture is that the translation into Eulerian quantities, such as the density field, invariably involves nonlinear combinations of $\bPsi$ even when only the linear equations of motion are considered.

\subsection{General Formalism}

While CDM particles by assumption experience only the gravitational force, baryons are subject to non-gravitational effects, such as Compton drag and pressure gradients. These effects can be summarized in the equations of motion of the fluid elements 
\begin{align}
    \ddot{\bPsi}_c + \mathcal{H}\dot{\bPsi}_c &= -\nabla_x\Phi(\textbf{q}+\bPsi_c) \nonumber \\
    \ddot{\bPsi}_b + \mathcal{H}\dot{\bPsi}_b &= -\nabla_x\Phi(\textbf{q}+\bPsi_b) + \textbf{F}_b(\textbf{q}+\bPsi_b),
    \label{eqn:species_eom}
\end{align}
where overdots signify derivatives with respect to $\tau$, $\mathcal{H}=d\ln a/d\tau$ is the conformal Hubble parameter, $\vb{F}_b$ is the non-gravitational force per unit mass felt by baryons, and $\Phi$ is the gravitational potential at Eulerian position $x$ satisfying Poisson's equation
\begin{equation}
    \nabla^2_x \Phi(\vb{x},\tau) = \frac{3}{2} \Omega_m(\tau) \mathcal{H}^2(\tau) \delta_m(\vb{x},\tau),
\end{equation}
where $\Omega_m$ is the total matter mass density and $\delta_m$ is the total matter overdensity (see below).

At the linear level, there is no difference between the Eulerian and Lagrangian positions in the above equations of motion, and we will neglect this distinction in the rest of this section unless otherwise stated. Indeed, taking the divergence of Equation~\ref{eqn:species_eom} in the linear limit ($x_\sigma \approx q$) directly yields the Euler equation when we map overdensities to displacements and velocities to their derivatives: 
\begin{equation}
    \delta_\sigma(\textbf{x}_\sigma) \leftrightarrow -\nabla \cdot \bPsi_\sigma(\textbf{q})\quad , \quad \textbf{v}_\sigma(x_\sigma) \leftrightarrow \dot{\bPsi}_\sigma(\textbf{q}).
    \label{eqn:correspondence}
\end{equation}
Note that the first mapping is correct only to linear order, while the second one is exact if the full $\mathbf{x}(\mathbf{q})$ is used. Assuming this translation, the solutions to the Lagrangian equations of motion as described below are essentially identical to those extracted from Boltzmann codes such as CAMB \cite{CAMB} or CLASS \cite{CLASS}, provided one chooses post-recombination initial conditions for the Lagrangian displacements.

To solve Equation~\ref{eqn:species_eom} in the linear limit, it is convenient to rewrite the baryonic and CDM displacements in terms of a mass-weighted matter component ($\bPsi_m = w_c \bPsi_c + w_b \bPsi_b$), which sources the gravitational potential, and a relative component that characterizes the differential flows between baryons and CDM ($\bPsi_r = \bPsi_b - \bPsi_c$), where we have defined the mass fractions of each species, $w_\sigma = \rho_\sigma/\rho_m$. These are related to the Eulerian quantities $\delta_m = w_b \delta_b + w_c \delta_c$ and $v_r = v_b - v_c$ by $\delta_a = -\nabla \cdot \bPsi_a$ and $v_a = \dot{\bPsi}_a$, where $a = \{m,r\}$, again at the linear level. The equations of motion in terms of these components are 
\begin{subequations}
\begin{align}
    \ddot{\bPsi}_m + \mathcal{H}\dot{\bPsi}_m &= -\nabla\Phi + w_b \vb{F}_b \label{eqn:m_eom} \\
    \ddot{\bPsi}_r + \mathcal{H}\dot{\bPsi}_r &= \vb{F}_b.
    \label{eqn:r_eom}
\end{align}
\label{eqn:mr_eom}
\end{subequations}
If in addition non-gravitational forces are negligible, the matter and relative components decouple, such that Equation~\ref{eqn:m_eom} can be solved as
\begin{equation}
    \bPsi_m(\tau) = -\textbf{m}_{+} D_{+}(\tau) + \textbf{m}_{-} D_{-}(\tau)
    \approx -\vb{m}_{+}D_{+}(\tau) \quad ,
    \label{eqn:matter_sol}
\end{equation}
where $D_{+}$ is the usual linear-theory growth factor. In the last step we have neglected the decaying mode, $\textbf{m}_{-}$, since it is a tiny fraction of the total displacement at all redshifts of interest. For non-gravitational forces, like Compton drag or pressure gradients, direct integration of the linear equations of motion show that the non-gravitational terms make a negligible contribution to the matter component $\bPsi_m$, such that the transfer function at redshifts below $z = 6$ agree with the linear solution in Equation~\ref{eqn:matter_sol} to within $0.2\%$, with even better agreement at the lower redshifts of interest in this paper. In the above we have included a minus sign for convenience such that $\delta_{m,0} = \nabla \cdot \textbf{m}_{+}$.

We end this subsection by discussing the full solution of the relative displacement when $\textbf{F}_b = \textbf{F}_b(\tau)$ is independent of $\bPsi_r$. In this case Equation~\ref{eqn:r_eom} is linear and first order in $\dot{\bPsi}_r$ and can be solved as:
\begin{equation}
    \dot{\bPsi}_r(\tau) = \textbf{v}_r(\tau_i)\, \Big( \frac{a_i}{a} \Big) + \frac{1}{a} \int_{\tau_i}^{\tau} d\tau'\ a(\tau') \textbf{F}_b(\tau'),
    \label{eqn:source_sol}
\end{equation}
where we have set the boundary conditions at initial time $\tau_i$ assuming the non-gravitational effects encoded in $F_b$ do not turn on until $\tau > \tau_i$. Equation~\ref{eqn:source_sol} turns out to be an excellent approximation for the large-scale Compton drag electrons experience in the reionization era due to their relative motion with respect to the CMB rest frame,  $\vb{F}_b = - n_e \sigma_T(\rho_\gamma/\rho_b)a \vb{v}_b$, where $\sigma_T$ is the Thompson scattering cross section, $\rho_\gamma$ is the photon energy density and $n_e$ the free electron number density. \eq{eqn:source_sol} also applies baryonic pressure forces on small scales $\vb{F}_b \propto - \nabla \delta_b$--- in both cases the total-matter component may be substituted for the baryonic component (i.e. $\delta_b \approx \delta_m$) at the sub-percent level \cite{Schmidt17}. In the case of the large-scale Compton drag, assuming $\vb{v}_b\simeq\vb{v}_m$ yields 
\begin{equation}
    \dot{\bPsi}_r(\tau) = \dot{\bPsi}_{r}(\tau_i) \frac{a(\tau_i)}{a(\tau)} + \Bigg[ \frac{1}{a} \int_{\ln(a(\tau_i))}^{\ln(a(\tau))} d\ln(a') \Bigg( n_e(a') \sigma_T \frac{\rho_\gamma(a')}{\rho_b(a')} \Bigg) \, \frac{f(a')D_{+}(a')}{a'^2} \Bigg] \frac{\bPsi_{m}(\tau_i)}{D_{+}(\tau_i)}\;,
    \label{eqn:compton_coeff}
\end{equation}
with $f=d D_{+}/d\ln(a) $ the linear theory growth factor. The Compton drag thus induces a mixing between the matter and relative components through a numerical prefactor dependent only on the linear growth factor $D_{+}$ and reionization history via $n_e$. Finally, we can integrate \ref{eqn:compton_coeff} to yield
\begin{equation}
    \bPsi_r(\tau) = - \vb{r}_{+} + \vb{r}_{-} D_r(\tau,\tau_i) + \vb{m}_{+} D_{\rm CD}(\tau,\tau_i), \qquad D_r(\tau,\tau_i) = \int^{\tau}_{\tau_i} \frac{H_0 d\tau'}{a(\tau')}
\end{equation}
where we can identify $\bPsi_r(\tau_i) = - \vb{r}_{+}$, $a(\tau_i) \vb{v}_r = H_0 \vb{r}_{-}$, and the Compton-drag kernel $D_{\rm CD}$ is defined as the conformal time integral of the square-bracketed function in \ref{eqn:compton_coeff}. The linear solutions to both the total-matter and relative components are thus wholly specified by the three modes $\vb{m}_{+}$ and $\vb{r}_\pm$. Jeans instabilities and baryonic pressure forces affect much smaller scales and won't be further discussed in the remainder of this work.

\begin{figure}
    \centering
    \includegraphics[width=\textwidth]{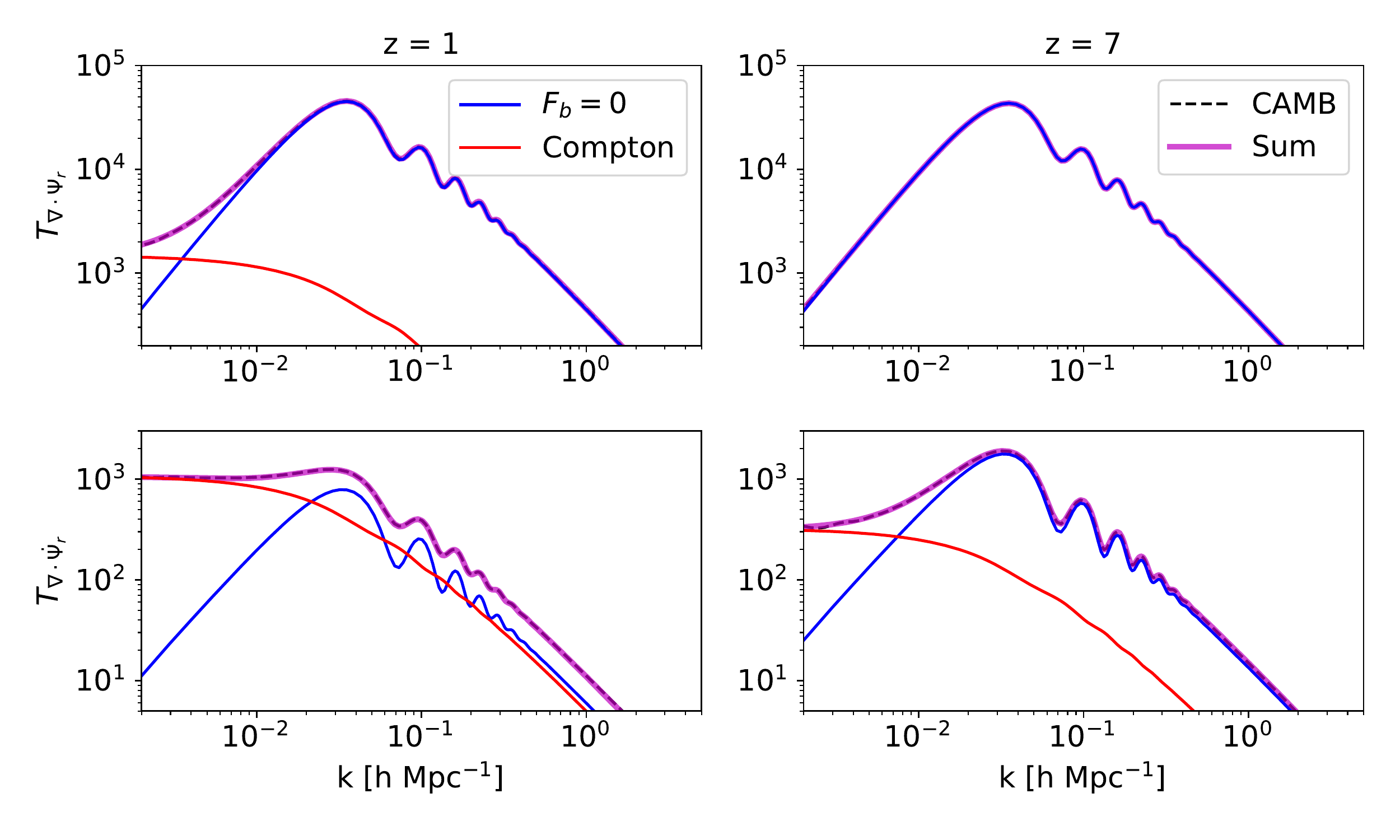}
    \caption{Transfer functions for the relative component from Equation~\ref{eqn:psidot_transfer} at $z = 1$ (left column) and $z = 7$ (right column). These transfer functions solve Equation~\ref{eqn:source_sol}.  The top row shows the transfer functions for $\nabla\cdot\bPsi_r$, i.e.~the relative density.  The bottom row shows the transfer functions for $\nabla\cdot{\dot{\bPsi}}_r$, i.e.~the relative velocity divergence. The free-falling ($F_b = 0$) and Compton drag contributions are shown separately, the effect of Compton drag on the relative velocity is immediately apparent even right after reionization ($z_{\rm re} = 7.90$) at $z = 7$, whereas the relative displacement is dominated by the $F_b = 0$ contribution at all but the largest scales shown. Unlike the Compton contribution, which is flat at large scales, the primordial ($F_b = 0$) contributions fall off as $k^2$ towards low wavenumbers, reflecting the origin of relative perturbations in pre-recombination baryonic pressure forces. At low redshifts, the solutions to the Lagrangian equations of motion, with initial conditions set at $z_i = 20$, are in excellent quantitative agreement with the results from CAMB (black dashed lines, barely visible on the plot as they lie below the purple lines).}
    \label{fig:transfer_func_sols}
\end{figure}

\subsection{Initial conditions and transfer functions}

The linear evolution of the density and velocity contrasts can be easily written in terms the CDM and baryon linear transfer functions (output from, e.g.~CAMB) as
\begin{equation}
    T_{\delta_r}(k) \equiv T_{\delta_b}(k) - T_{\delta_c}(k)
    \quad\mathrm{and}\quad
    T_{\theta_r}(k) \equiv T_{\theta_b}(k) - T_{\theta_c}(k)
\end{equation}
where $\theta_{b,c}(k) \equiv  -\dot{\delta}_{b,c}(k)$. It is worth noticing that while the velocity field is gauge dependent, velocity differences are not.
The transfer function for $\nabla \cdot\mathbf{m}_{+}$ is simply the present-day matter transfer function $T_m$ and we can furthermore define
\begin{align}
T_{\nabla \cdot\mathbf{r}_{+}}(k) &\equiv T_b(k,z_i) - T_c(k,z_i) \nonumber \\
    T_{\nabla \cdot\mathbf{r}_{-}}(k) &\equiv [(1+z_i)H_0]^{-1} \Big( T_{\theta_b}(k,z_i) -T_{\theta_c}(k,z_i)\Big).
    \label{eq:Tpsi}
\end{align}
These three functions specify the solution for the $\bPsi_m$, $\bPsi_r$ and $\dot{\bPsi}_r$  at any $z<z_i$. The choice of $z_i$ is somewhat arbitrary  but choosing redshifts before the onset of reionization has the advantage of separating the effects of gravity from Compton drag. This choice also justifies the normalization in \eq{eq:Tpsi}, since $\vb{r}_-$ is independent of redshift. In the remainder of the paper we assume $z_i=20$.

In addition to the above, we will show below that calculating the power spectrum at some redshift $z$ in the Lagrangian picture requires linear-theory spectra of the relative displacement at that redshift, which will typically include corrections from Compton drag. These can be calculated via Equations~\ref{eqn:matter_sol} and \ref{eqn:source_sol} to give
\begin{equation}
    T_{\nabla \cdot \bPsi_r}(k,z) = T_{\nabla \cdot\mathbf{r}_{+}}(k) + D_r(z,z_i) T_{\nabla \cdot\mathbf{r}_{-}}(k) + D_{\rm CD}(z,z_i) T_{\nabla \cdot m_{+}}(k).
    \label{eqn:psidot_transfer}
\end{equation}
Sample solutions of the equation of motion in \eq{eqn:mr_eom} when $F_b$ is given by Compton drag with the CMB are shown in Figure~\ref{fig:transfer_func_sols}. After reionization most of large scale power in the relative velocity transfer function, $T_{\nabla\cdot \dot{\bPsi}_r}$, is provided by the Compton drag, which in turn affects the evolution of the relative baryon-dark matter density  at large scales (see top panels in \fig{fig:transfer_func_sols}). \fig{fig:transfer_func_sols} also justifies the approximations we used to compute the drag forces, as one can see by the excellent agreement with the full CAMB output. Other non-gravitational effects like pressure terms (Jeans instability) and radiative transfer effects \cite{Pontzen,Gontcho,Upton,Cabass}, can be written in a similar form.

\begin{figure}
    \centering
    \includegraphics[width=\textwidth]{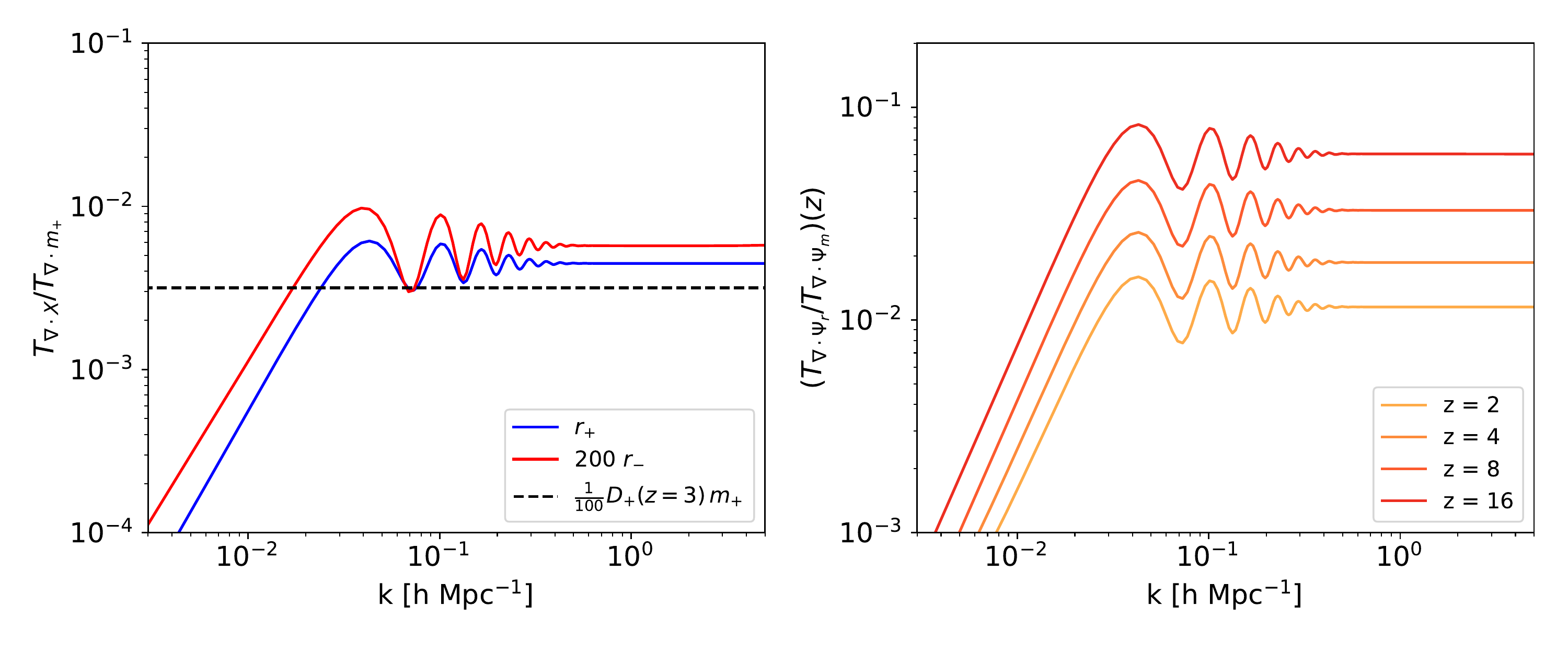}
    \caption{Relative to total-matter-component transfer function ratios. (Left) Transfer function ratios between the initial fields $m_{+}$ and $r_\pm$ defined at $z = 20$. The so-normalized constant $r_{+}$, which roughly corresponds to the relative overdensity mode, is a percent level contribution relative to the total-matter growing mode $m_{+}$. The decaying mode $r_{-}$, which corresponds roughly to the relative velocity, enters at significantly below the percent level. Note however that our definition somewhat exaggerates its smallness by ``redshifting'' it to $z = 0$. The equivalent ratio for one percent of the growing mode at $z = 3$ is plotted for comparison in black. (Right) Transfer function ratios between the evolved relative and total matter displacements at redshifts $z = 2-6$. While the relative displacement is a percent level effect at low redshifts ($z = 2$), it enters at close to the ten percent level at higher redshifts ($z=16$).
    }
    \label{fig:tf_ratios}
\end{figure}

Ratios of the transfer functions to the total matter one are shown in Figure~\ref{fig:tf_ratios}. We notice that the relative density perturbation is much larger than the relative velocity one, by a factor of a hundred at least, and the two relative components have the same behavior with wave-number $k$ at small and large scales. Nonetheless $\vb{r}_+$ and $\vb{r}_-$ have significant differences in shape around the BAO scales and therefore will have to be treated separately from the point of view of the galaxy bias expansion.

\section{Lagrangian Bias in the Two-Fluid Dynamics}
\label{sec:bias}

In the Lagrangian approach, galaxy bias is assumed to arise as the response of the overdensity of galaxies, or the precursors thereof, to the variation of the initial conditions encoded in the fields $\{\bPsi_{\sigma}(\textbf{q})\}$ of the various species, and then transported via advection to their present-day positions $\bx(q,t) = \bq + \bPsi_g(\textbf{q},t)$.
Thus, when computing the density of a biased tracer the number-conservation Equation~\ref{eqn:dens_eq} is modified to

\begin{equation}
    1 + \delta_g(\textbf{x},\tau) = \int d^3q \; F_g[\vb{q}|\,\{ \bPsi_{\sigma}(\textbf{q})\}] \, \delta_D\left[\textbf{x} - \textbf{q} - \bPsi_g(\textbf{q},\tau)\right].
    \label{eqn:tracer_dens}
\end{equation}

The standard picture of (local) Lagrangian bias, outlined above, has been developed in the 1-fluid case by many authors, see for example \cite{Cat98, Mat08a, Mat08b,Bal15a,Desjacques10, ESP, Modi,VlaCasWhi16,Schmittfull18} and \cite{Desjacques16} for a recent review on galaxy bias. 
In this section our focus will be on extending these arguments to the case of multiple fluids, and in particular to the two-fluid case. In the presence of two fluids, the form of Equation~\ref{eqn:tracer_dens} raises two questions: (1) the form of the response $F_g$ and (2) whether biased tracers follow the dark matter, baryons, or a combination thereof. We address these in turn.

\subsection{Bias Expansion}

The initial tracer overdensity, $F_g[\vb{q}|\,\{\bPsi_{\sigma}(\textbf{q})\}]$, is defined to be a functional encoding the physics of gravitational collapse and galaxy formation at some Lagrangian position $\mathbf{q}$. Since the galaxy density field is a scalar quantity under rotations, $F_g$ will also be a scalar. We will assume this functional is local, in the sense gravitational collapse depends only on the value of the fields within a characteristic patch of size $R_h$, which then flows coherently on large scales with $\bPsi_g$ \cite{Desjacques16}. In the fluid limit, these conditions imply that the system is wholly specified -- albeit in some complex, nonlinear way -- by the species overdensities, $\delta_\sigma(\textbf{q})$, velocities, $\textbf{v}_\sigma(\textbf{q})$, and the gravitational potential\footnote{The gravitational potential $\Phi$, while not independent of $\delta_m$, depends on the total matter density in a very non-local way. To make our bias expansion local, and be able to truncate the derivative expansion at a reasonable order, we thus include it as a standalone quantity here.}, $\Phi(\textbf{q})$, at some initial time $\tau_i$.  The condition that $F_g$ is local -- or rather, nonlocal with width $R_h$ -- can be equivalently (but more conveniently) expressed by requiring $F_g$ depend only on the initial fields and their spatial derivatives, with $n^{\rm th}$ derivatives suppressed by $n$ powers of $R_h$ \cite{Desjacques16}.

In addition to the assumption of locality, the form of $F_g$ is strongly restricted by various symmetries. General relativity requires that all physical quantities be diffeomorphism invariant, which in our case reduces to rotational invariance and invariance under generalized Galilean transformations \cite{Horn}:
\begin{equation}
  \vb{q} \rightarrow \vb{q} \quad , \quad \bPsi_\sigma \rightarrow \bPsi_\sigma + \textbf{n}(\tau)
  \quad , \quad 
  \Phi \rightarrow \Phi \rightarrow \Phi - \textbf{x} \cdot (\ddot{\textbf{n}} + \mathcal{H} \dot{\textbf{n}}) \quad ,
    \label{eqn:galilean}
\end{equation}
where $\mathbf{n}$ are time-dependent but spatially constant vector fields. 

Rotational invariance simply requires that only contracted scalar quantities enter the bias; the restrictions placed on the bias expansion by general Galilean invariance are more subtle, and it is here that the two-fluid case diverges from the single-fluid case. Under this symmetry, densities remain unchanged--- for instance that at first order $\delta_\sigma(\vb{q})  = - \nabla\cdot \bPsi_\sigma(\vb{q})$--- while velocities get boosted by a spatially constant amount (leaving $\partial \textbf{v}$ invariant) and the gravitational potential changes in a spatially linear way (leaving $\partial \partial \Phi$ invariant). In the single-fluid regime, where only one set of densities and velocities exist, this directly implies that velocities can only enter with at least one spatial derivative, and the gravitational potential can only enter as second (spatial) derivatives and beyond. The single-fluid overdensity, which is unchanged under the transformation, can enter at any order. 

The presence of two or multiple fluids relaxes some of the above restrictions. In particular, since all particle velocities are boosted by the same amount ($\textbf{n}'$) under a Galilean transformation, the relative velocity $\textbf{v}_r = \textbf{v}_b - \textbf{v}_c$ remains invariant and can thus enter the bias expansion at zeroth order in derivatives. The total matter velocity, $\textbf{v}_m$, on the other hand, is boosted and can thus still only enter at the derivative level. These two quantities form an equivalent basis to the individual species velocities and there is no loss of generality in defining the bias expansion in terms of them. We may similarly write terms involving species densities, which can enter separately, in the total matter and relative density basis. In general relativity the gravitational potential is unaffected by the number of species as a consequence of the equivalence principle, \ie gravitational interactions are universal. The full set of physical fields that can enter $F_g$ in the two fluid case is then
\begin{equation}
    F_g = F_g \left[ \delta_{\sigma}, \textbf{v}_{\sigma}, \Phi \right] \equiv F_g \left[ \delta_{m}, \delta_{r}, \partial \textbf{v}_{m} ,\textbf{v}_{r}, \partial \partial \Phi ,\cdots\right],
\end{equation}
where the dots stand for higher derivative operators.
To first order in the fields we can therefore write\footnote{A list of bias terms up to second order is given in Appendix~\ref{sec:second_order}.}
\begin{equation}
    1 + \delta_{g}(\vb{q}) = 1 + b_1 \delta_{m} + b_r \delta_{r} + b_\theta \theta_{r} + \cdots
\label{eqn:bias_conventional}
\end{equation}
which is similar to the Eulerian linear theory expression in \cite{Schmidt16}. This is not surprising, since at first order $\vb{q}\simeq \vb{x}$, however we will see below that differential advection can introduce further terms degenerate with the initial Lagrangian bias terms above, such that the Eulerian relative-component bias will in general be a combination of these terms.

Finally, since $F_g$ is defined as a functional on the initial conditions which can be chosen to be sufficiently early that they lie deep in the linear-theory regime, its form can be further simplified and expressed purely in terms of the initial modes $\textbf{m}_{+}$ and $\textbf{r}_\pm$. In the single fluid case, this restriction leads to the simplification that all bias terms can be written in terms of spatial derivatives of the total matter displacement $\textbf{m}_{+} \sim \bPsi_m$; this is a direct consequence that, up to time-dependent constant factors, $\delta_m \sim \partial \bPsi_m$, $\textbf{v}_m \sim \bPsi_m$ and $\partial \partial \Phi \sim \partial \bPsi$ in linear theory. In the two-fluid case these terms must be supplemented by those involving the relative modes. Specifically, including the $\textbf{v}_{r}$ dependence requires the inclusion of terms proportional to $\textbf{r}_{-}$ and including $\delta_{r}$ dependence similarly requires terms proportional to $\nabla \cdot \textbf{r}_{+}$. Equation~\ref{eqn:bias_conventional} can thus be re-expressed as:
\begin{align}
     F_g (\vb{q}) = b_1 \delta_m +  b_{{+}} \nabla \cdot \textbf{r}_{+} + b_{-} \nabla \cdot \textbf{r}_{-} +\,...
    \label{eqn:bias_exp}
\end{align}
 We therefore have a direct correspondence in the bias expansion between the initial modes expressed in Eulerian and Lagrangian space. Notice that the bias expansion defined above is complete, in the sense that it contains all possible operators compatible with the symmetries of the problem. In particular, while $\vb{r}_\pm$ are defined at a particular initial redshift $z_i$, in the linear regime this dependence amounts to a simple linear transformation and can be absorbed into the definition of the bias parameters (Appendix~\ref{sec:running_bias}). 
 
 Finally, an additional complication arises when halo formation is affected by Compton drag. As pointed out by \cite{Schmidt17}, by picking out the local CMB rest frame such that the drag force $\propto \vb{v}_b$, we lose the gauge redundancy of Galilean transformations. This will in general produce heretofore forbidden terms such as those proportional to the matter-component velocity $\vb{v}_m$. However, the terms thus generated are required by rotational invariance to enter at second order and beyond. For the remainder of this paper we will thus neglect these contributions, which are subdominant to the already sub-percent level contributions we study.

Whereas there exists quite a large literature on measuring and predicting, using approximate physical models, the value of the bias parameters in one-fluid scenarios, less attention has been devoted to the multi-fluid case. From an effective field theory perspective the dimensionless parameters should be of order unity, but in reality the actual value of the bias parameters is tracer-dependent and can be quite a bit larger or smaller.
In this work we  will assume, unless otherwise noted, that typical values are given by $b_+\simeq1$ and $b_-\simeq6.8$  derived in \cite{Schmidt16} using a spherical collapse model. These numbers are consistent with the non-detection of relative bias effects in BOSS DR12 by ref.~\cite{Beutler2017}, who find e.g.\ $b_+ = -1.0 \pm 2.5$ to within one sigma when fitting for $b_+$, $b_-$ and $c_-$ (Section~\ref{ssec:higher_order_bias}) across all redshift bins, with large systematic biases measured in dark-matter only simulations that had to be subtracted. 

\subsection{Modifications to Tracer Advection}

Once the initial, biased tracer overdensity is set, the overdensity at later times is set by the tracer ``fluid'' advecting from initial (Lagrangian) $q$ to final positions $q + \bPsi_g$ along trajectories described by the tracer equation of motion
\begin{equation}
    \ddot{\bPsi}_g + \mathcal{H} \dot{\bPsi}_g = - \nabla \Phi + \vb{F}_{b,\,g},
    \label{eqn:tracer_eom}
\end{equation}
where we have included a non-gravitational term, $F_{b,\,g}$, to account for the possibility that tracers feel non-gravitational forces. Such non-gravitational contributions may arise, for example, from the Compton drag on the baryonic component of galaxies, or from various galaxy formation processes. Since such contributions are always local in space and time, we will assume the above equation satisfies the same symmetries of \eq{eqn:galilean}, \ie the force acting on galaxies depends only on density fields and velocity gradients.

Equation~\ref{eqn:tracer_eom} can be solved by subtracting the equation of motion of the total matter displacement (Equation~\ref{eqn:m_eom}) and defining $\bPsi_{r,g} = \bPsi_g - \bPsi_m$. Neglecting the baryonic contributions such that the tracers' dynamics are governed only by gravity, and assuming that the initial tracer displacements are a weighted average of the baryonic and CDM displacements, i.e. $\bPsi_{g,i} = \bPsi_{m,i} + f_g \bPsi_{r,i}$, this immediately yields the time evolution
\begin{equation}
    \bPsi_{g}(\tau) = \bPsi_{m}(\tau) + f_g [\bPsi_{r}(\tau)]_{\rm CD = 0},
    \label{eqn:tracer_sol}
\end{equation}
where the relative displacement is evaluated assuming zero Compton drag. Note that if we assume that the tracer field is made of objects composed of the same mass fractions of baryons and CDM as the total matter content of the universe, i.e. with $f_g = 0$, Equation~\ref{eqn:tracer_sol} reduces to the trajectory of the matter component. Similarly, objects composed purely of baryons or the CDM will (at the linear level) follow the baryon or CDM displacements, respectively. 

We can alternatively think of \eq{eqn:tracer_sol} as a bias expansion of the galaxy displacements in terms of the underlying fields, since $\bPsi_m$ and $\bPsi_r$ are the only two linear operators allowed by symmetries at lowest order in spatial derivatives. If the tracer flow is purely gravitational, as assumed above, the equivalence principle further restricts the coefficient of the total matter displacement -- which encapsulates the motion due to the gravitational potential -- to be exactly 1 at all times. However, this restriction can be broken by baryonic contributions ($\propto \vb{F}_{b,\,g}$) such as the Compton drag. As seen in the second term on the right hand side of Equation~\ref{eqn:compton_coeff}, the acceleration due to Compton drag generates displacements proportional to $\bPsi_m$; this contribution, on top of the aforementioned gravitational displacements, can lead to an expansion $\bPsi_g = (1 + \alpha_{\rm CD}) \bPsi_m + f_g \bPsi_r + ...$ for some nonzero coefficient $\alpha_{\rm CD}$ due to Compton drag, where the total-matter coefficient deviates from unity. Consequences of this modified expansion for the power spectrum are considered at the end of Section~\ref{ssec:linear_spectra} and in Figure~\ref{fig:compton_pk}. Other baryonic forces, such as pressure forces at small scales, can similarly be included as further terms ($\bPsi_g \ni c_s^2 \nabla \delta_b$) in this expansion.

\section{Galaxy Power Spectra in the Zeldovich Approximation}
\label{sec:zeldovich}

\subsection{Analytic Form}
\label{ssec:linear_spectra}

From Equation~\ref{eqn:tracer_dens}, the power spectrum at redshifts $z$ for a biased tracer can be computed as
\begin{equation}
    P_{gg}(k,z) = \int d^3q \; e^{i \vb{k}\cdot \vb{q}}  \left\langle F_{g}[\vb{q}_1] F_{g}[\vb{q}_2] \, e^{i k \cdot (\bPsi_g(\vb{q}_1,z) - \bPsi_g(\vb{q}_2,z))} \right\rangle_{q=|\vb{q}_2-\vb{q}_1|},
\label{eqn:power_spec}
\end{equation}
where the subscripts denote quantities evaluated at two points separated by $q$ in Lagrangian space. It is important to note that the bias functions $F_{g}$ are evaluated in terms of the linear modes $m_{+}, r_\pm$ defined at the initial redshift $z_i$. In the Zeldovich approximation displacements are solved to linear order but the full mapping between initial and final times is kept.  This amounts to keeping the displacement correlators exponentiated in what follows \cite{Mat08a}.
We will adopt the bias expansion in Equation~\ref{eqn:bias_exp}. We evaluate integrals involving $F_g$ by functional differentiation in the usual manner \cite{Mat08a,Mat08b,CLPT}: we include a term (e.g.~$\lambda X$) in the exponential for each argument, $X$, of $F_g$ and evaluate terms like $X^n$ via $\partial^n/\partial\lambda^n$ of $\exp[\lambda X]$.

\begin{figure}
    \centering
    \includegraphics[width=\textwidth]{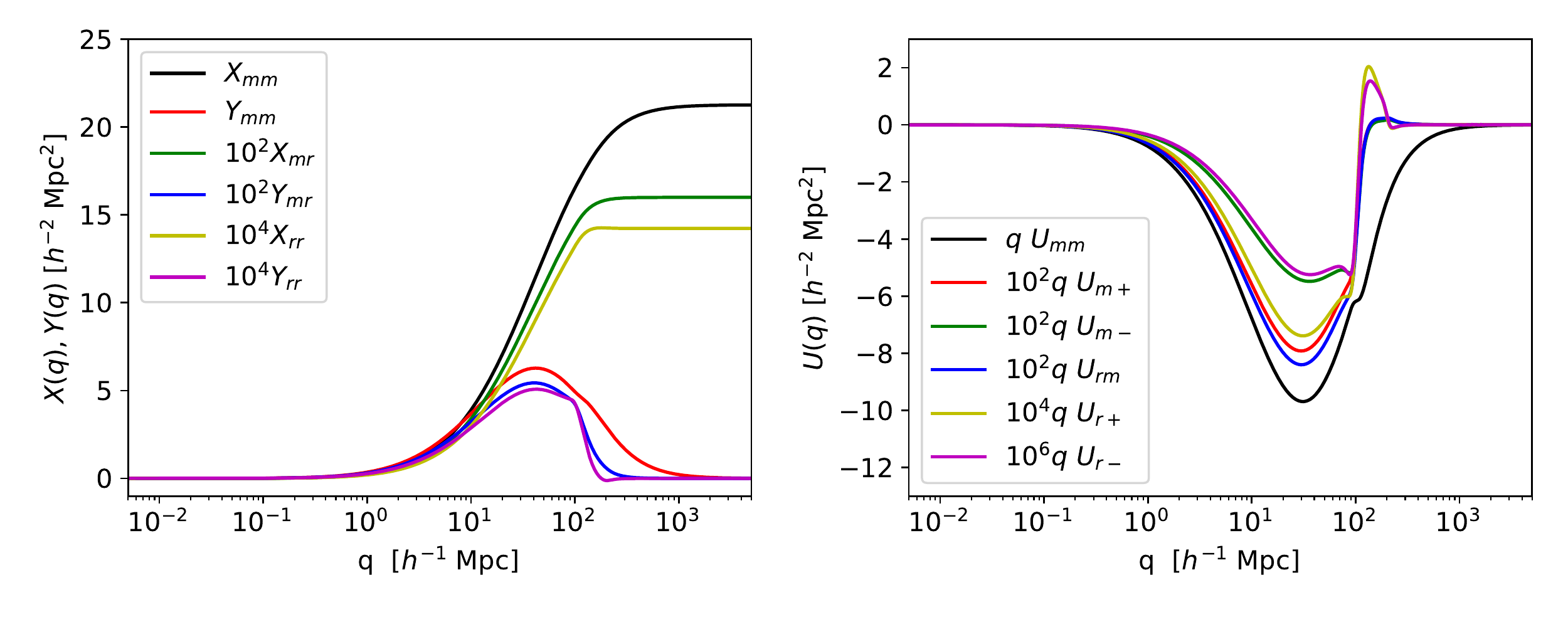}
    \caption{Correlation functions entering the galaxy power spectrum in \eq{eqn:pk_zeldovich} at $z=1.2$. Left panel: the displacement auto- and cross-correlation functions between the different components. Right panel: bias-weighted, displacement correlation functions. Correlation functions involving the relative component exhibit abrupt features around $q\sim 10^2\ h^{-1}$ Mpc, reflecting the baryon acoustic oscillation scale.
    }
    \label{fig:q-func}
\end{figure}

Under the above assumptions our task reduces to evaluating
\begin{equation}
  e^{i \mathcal{M}} \equiv   \left\langle\exp\left(i \vb{k}\cdot\vb{\Delta}_g(z) + \lambda_{\delta_{m,1}}\delta_{m,1}+ \lambda_{+,1} \nabla \vb{r}_{+,1} + \lambda_{-,1 }\nabla \vb{r}_{-,1}
    + (1\leftrightarrow2) \,\right)\right\rangle
\end{equation}
with numerical subscripts referring to Lagrangian coordinates, $q_1$ and $q_2$, and 
\begin{equation}
\vb{\Delta}_g = \bPsi_{g,1} - \bPsi_{g,2} =     \bPsi_{m,1} - \bPsi_{m,2} + f_g (\bPsi_{r,1} - \bPsi_{r,2}) \equiv \vb{\Delta}_m +f_g \vb{\Delta}_r
\end{equation}
The function $e^{i\mathcal{M}}$ can be evaluated using the cumulant theorem as the exponential of the connected components.     The Zeldovich approximation assumes linear dynamics, such that only quadratic terms survive
\begin{align}
e^{i \mathcal{M}} =     \exp\left\{\vphantom{\int} \right. & -\frac{1}{2} k_i k_j A^{mm}_{ij} -  f_g k_i k_j A^{rm}_{ij} - \frac{f_g^2}{2} k_i k_j A^{rr}_{ij}  \nonumber\\
    &+ ik \cdot \big((\lambda_{\delta_{m,1}}+\lambda_{\delta_{m,2}}) (U_{mm} + f_g U_{rm}) \nonumber \\
    &+ (\lambda_{+,1}+\lambda_{+,2}) (U_{m+} + f_g U_{r+}) + (\lambda_{-,1}+\lambda_{-,2}) (U_{m-} + f_g U_{r-}) \big) \nonumber \\
    &+  (\, \lambda_{\delta_{m,1}}\lambda_{+,2} + (1\leftrightarrow2)\,) \; \xi_{\delta_m \nabla \vb{r}_+} + (\delta_m, \nabla \vb{r}_-) + (\delta_m, \delta_m)  \nonumber \\
    &+ \left. (\nabla \vb{r}_+, \nabla \vb{r}_+) + (\nabla \vb{r}_+,\nabla \vb{r}_-) + (\nabla \vb{r}_-, \nabla \vb{r}_-)
    \vphantom{\int}\right\} \quad ,
    \label{eqn:second_cumulant}
\end{align}
where we have defined
\begin{equation}
    A^{ab}_{ij} = \left\langle \Delta^a_i(z) \Delta^b_j(z) \right\rangle, \quad U^{a\pm}_i = \left\langle \Delta^a_i(z) \nabla \cdot \vb{r}_\pm(\vb{q}_1)  \right\rangle, \quad \xi_{a b} = \left\langle a(\vb{q}_1)b(\vb{q}_2) \right\rangle
    \label{eqn:defs},
\end{equation}
noting that the $\Delta$'s carry an implicit redshift dependence while the other fields do not. For the total-matter component this redshift dependence is a direct growth factor scaling and we will for convenience take the linear field's value as evaluated at the observed redshift $\delta_m = - D_m(z) \nabla \cdot \vb{m}_+ $. The paired parentheses denote terms similar to the preceding except with the indicated pair of variables.  For example, in the third line
\begin{equation}
    (\delta_m, \nabla \vb{r}_-) \equiv   \left(\, \lambda_{\delta_{m,1}}\lambda_{-,2} + (1\leftrightarrow2)\,\right) \; \xi_{\delta_m \nabla \vb{r}_-}(q)
\end{equation}
and when the elements of a pair are repeated the term should be divided by a symmetry factor of two.  

\fig{fig:q-func} shows the different correlation functions entering the above calculation. Since the correlation function of the different displacements fields, $A^{ab}_{ij}(q)$, is a tensor, we can decompose it as $A^{ab}_{ij}(q) = X^{ab}(q) \delta^{K}_{ij} + Y^{ab}(q) \hat{q}_i\hat{q}_j$, and the functions $X(q)$'s and $Y(q)$'s are shown in the left panel of \fig{fig:q-func}. Clearly the galaxy displacements are dominated by the total matter component, with the relative terms contributing much less than a \% to the bulk flows. This fact will enable us to treat the terms proportional to $f_g$ perturbatively, as they will be much smaller than one for wavenumbers below the nonlinear scale defined by $k^2 \Sigma^2 \lesssim 1$, where the Zeldovich r.m.s. displacement is $\Sigma \propto X_{mm}(q \rightarrow \infty)$. 
The same conclusions apply for the bias weighted displacements $U(q)$'s, shown on the right hand panels in \fig{fig:q-func}, where $U_m(q)\gg U_\pm(q)$.

\begin{figure}
    \centering
    \includegraphics[width=\textwidth]{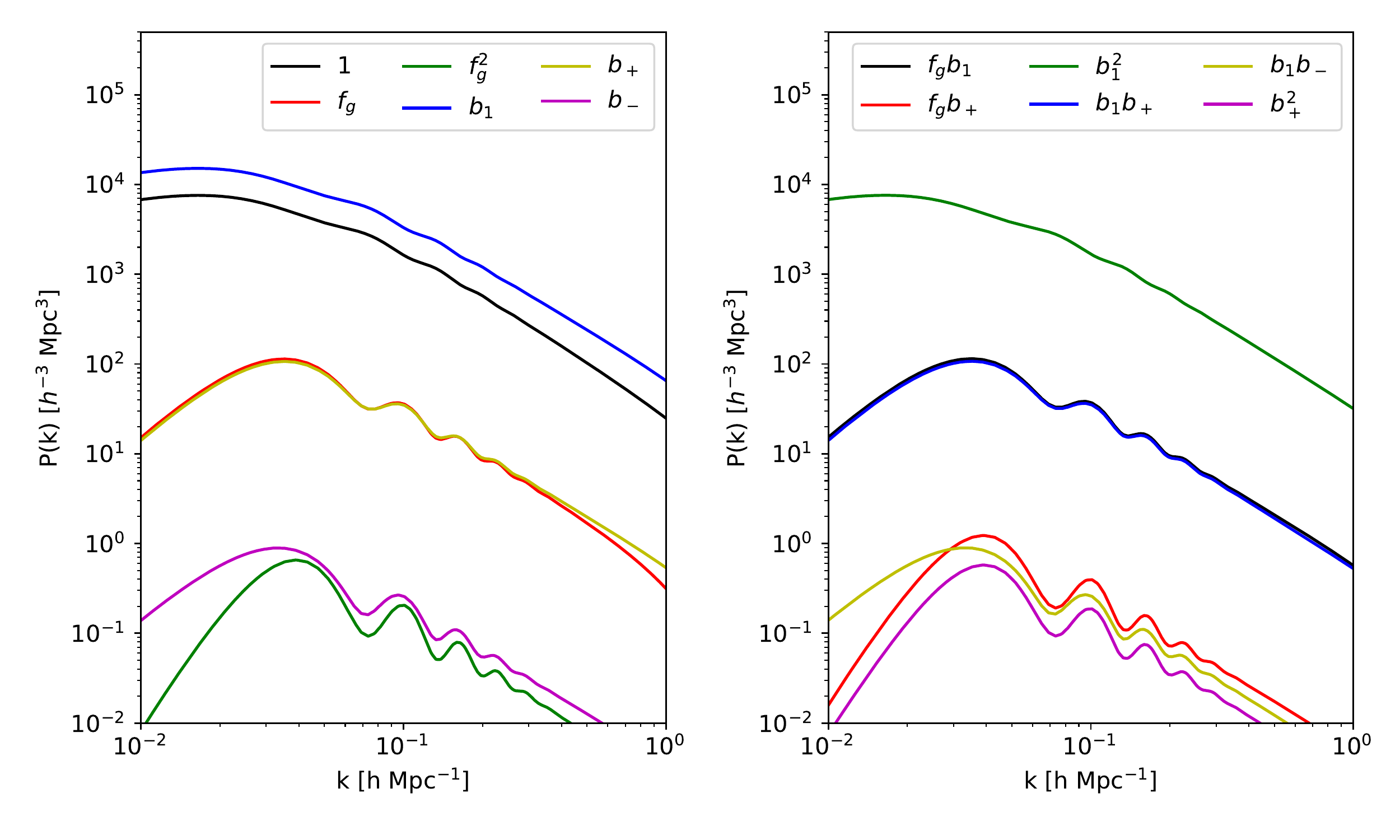}
    \caption{Different contributions to the galaxy power spectrum in the Zeldovich approximation, \eq{eqn:pk_zeldovich}, at $z=1.2$. Terms proportional to $b_+ b_-$, $f_g b_-$, and $b_-^2$ have been omitted as they are two orders of magnitude smaller than the smallest contributions shown. Many terms, such as those involving $f_g$ and $b_+$, are essentially degenerate.
    }
    \label{fig:Pks}
\end{figure}

Working to linear order in the power spectrum we then have that the galaxy-galaxy power spectrum is given by
\begin{eqnarray}
    P_{gg}(k) = \int d^3q\ e^{i \vb{k}\cdot\vb{q}}\,e^{-\frac{1}{2}k_i k_j A^{mm}_{ij}} &\Big[& 1 - f_g k_i k_j A^{rm}_{ij} - \frac{f_g^2}{2} k_i k_j A^{rr}_{ij} \nonumber \\
    &+& 2ik\cdot (b_1 U_{mm} + b_+ U_{m+} +  b_- U_{m-}) \nonumber \\
    &+& 2f_g ik\cdot (b_1 U_{rm} + b_+ U_{r+} +  b_- U_{r-}) \nonumber \\
    &+& b_m^2 \xi_{\delta_m \delta_m} + 2 b_m b_+ \xi_{\nabla \vb{r}_+ \delta_m} + 2 b_m b_- \xi_{\nabla \vb{r}_- \delta_m} \nonumber \\
    &+& b_+^2 \xi_{\nabla \vb{r}_+ \nabla \vb{r}_+} +  2 b_+ b_- \xi_{\nabla \vb{r}_+ \nabla \vb{r}_-} +  b_-^2 \xi_{\nabla \vb{r}_- \nabla \vb{r}_-} + \mathcal{O}(P_L^2) \Big].
    \label{eqn:pk_zeldovich}
\end{eqnarray}
\fig{fig:Pks} shows the different contributions to the galaxy power spectrum in the Zeldovich approximation at $z = 1.2$. The leading corrections to the total-matter power spectrum come at the roughly percent level from terms in Equation~\ref{eqn:pk_zeldovich} linear in $\vb{r}_+$, i.e. in $b_+$ and $f_g$. These contributions are essentially degenerate, with differences due to the dynamical evolution of $\bPsi_r$ in the $f_g$ term, as we will discuss in the next paragraph. Corrections quadratic in $\vb{r}_{+}$ or linear in $\vb{r}_{-}$ enter at roughly the same size four orders of magnitude below the total-matter contributions.

An interesting consequence of the advection of biased tracers with $|f_g| > 0$ is the appearance of relative bias terms even if none were present in the initial Lagrangian bias expansion. To see this, we can take the low-$k$ limit of \eq{eqn:pk_zeldovich}, neglecting for the moment non-gravitational contributions to $\bPsi_r(q)$, and obtain up to $\mathcal{O}(P(k))$
\begin{align}
    P_{gg}(k,z) = & (1+b_1)^2 \, P_{\delta_m \delta_m}(k) \nonumber \\
    &+   2 (1 + b_1)(b_+ + f_g)  P_{m \nabla \vb{r}_+}(k) +  2 (1 + b_1)(b_-+ f_g D_r(z)) \, P_{m \nabla \vb{r}_-}(k) \nonumber \\ &+ (b_+ + f_g)^2 P_{\nabla\vb{r}_+ \nabla\vb{r}_+}(k)+(b_- + f_g D_r(z))^2 P_{\nabla\vb{r}_- \nabla\vb{r}_-}(k) \nonumber \\&+ 2 (b_+ + f_g)(b_- + f_g D_r(z))P_{\nabla\vb{r}_+ \nabla\vb{r}_-}(k)\,.
\end{align}
We immediately recognize the familiar expression for the Eulerian linear bias, $b^E_1 = 1+b_1$, and that the relative density and velocity bias terms get renormalized by terms proportional to $f_g$. To make further contact with the existing literature employing the Eulerian formulation of the equations of motion \cite{Schmidt16,Schmidt17}, we can identify the relative baryon dark-matter density perturbation $\delta_{r}$ with the divergence of $\vb{r}_+$, $\delta_{r} \equiv \nabla \cdot \vb{r}_+$, and the relative baryon dark-matter velocity divergence $\theta_{r}$ with the divergence of $\vb{r}_-$,  $\theta_{r} \equiv (1+z)H_0\nabla \cdot \vb{r}_-$. This implies that the bias parameters in \cite{Schmidt16,Schmidt17} associated to the Eulerian fields are $b^E_{\delta_{r}} = b_+ + f_g$ and $b^E_{\theta_{r}} =(1+z)^{-1}H_0^{-1} (b_- + f_g D_r(z))$. Note that the referenced overdensities and velocities are those defined at the initial redshift $z_i$ so should not be directly substituted for their Eulerian counterparts; for more details about the mapping of bias parameters from some initial time $z_i$ to Eulerian coordinates see Appendix \ref{sec:running_bias}.

\begin{figure}
    \centering
    \includegraphics[width=\textwidth]{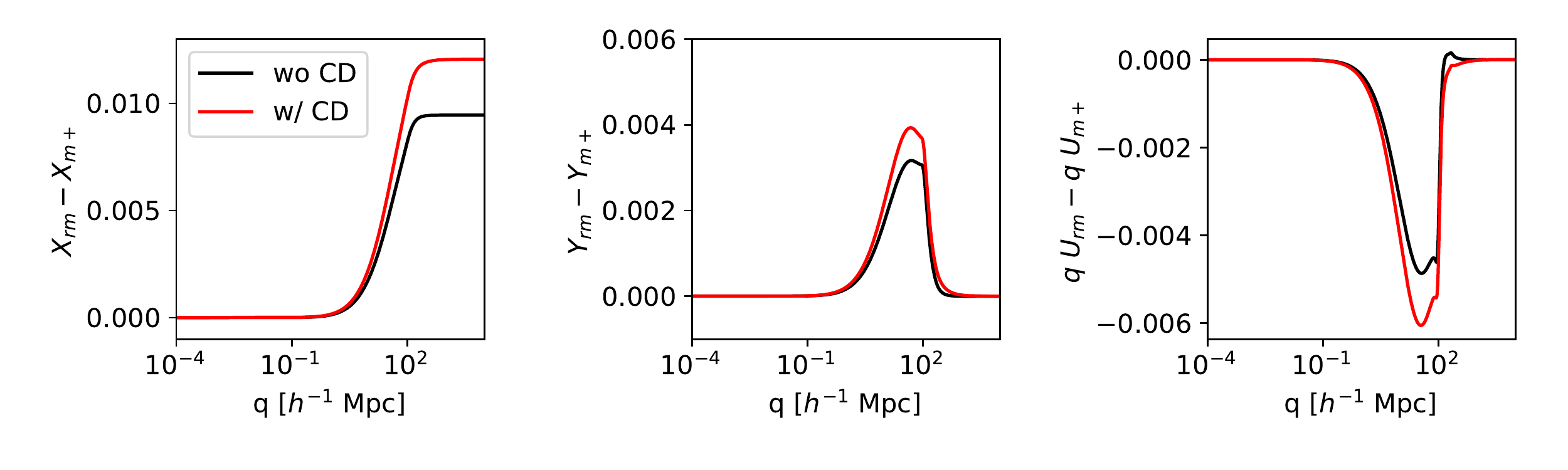}
    \caption{Comparison of two point functions with (red) and without (black) contributions from Compton drag. While the differences are small (c.f.\ Fig.~\ref{fig:q-func}), they are non-neglible at large scales. The contributions from $\vb{r}_{+}$ have been subtracted off for ease of comparison.
    }
    \label{fig:compton_2pt}
\end{figure}

\begin{figure}
    \centering
    \includegraphics[width=\textwidth]{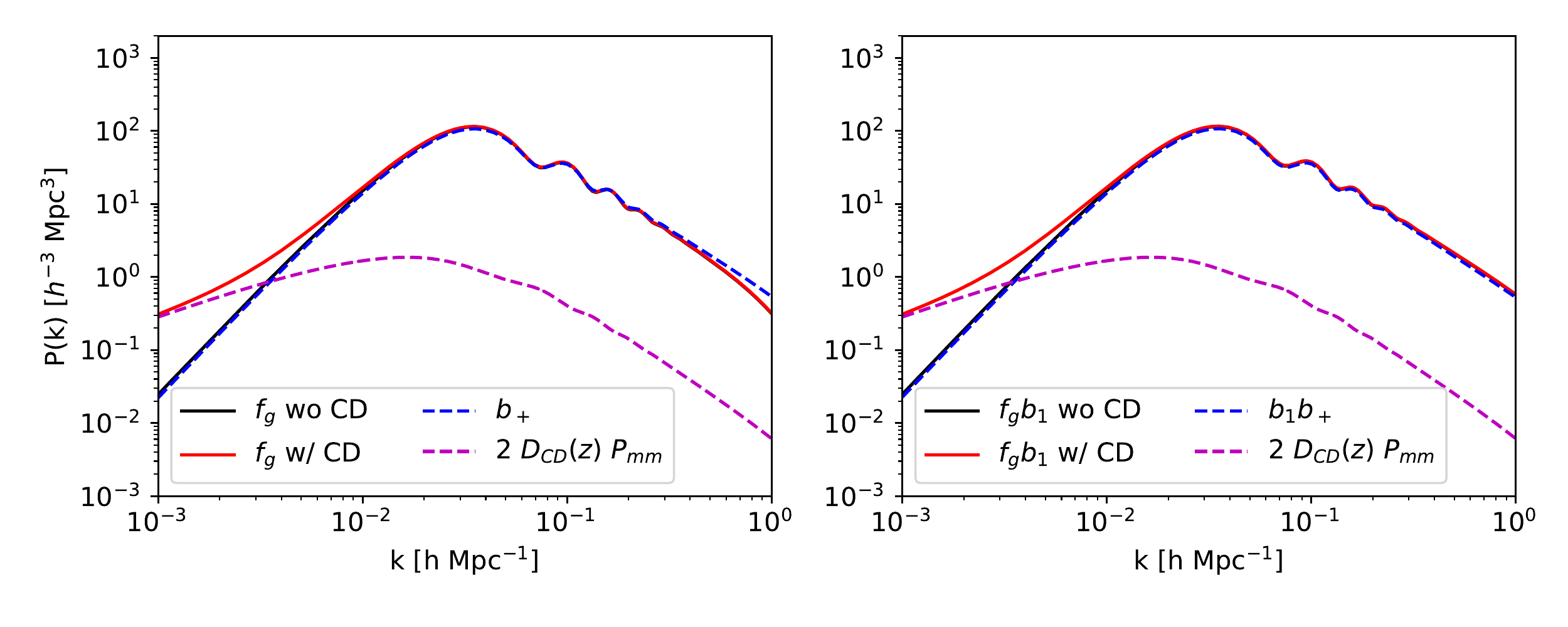}
    \caption{Comparison of terms involving $b_+$ (blue dashed) and $f_g$ with and without Compton drag (red and black). The two are largely degenerate in the latter case, but with Compton drag the $f_g$ terms are dominated by a contribution proportional to the total-matter power spectrum at large scales, which can alternatively be renormalized into the matter bias $b_1$, shown separately as a dashed magenta curve. The left panel shows contributions due to contracting the relative components ($f_g \bPsi_r$ or $b_+ \nabla \cdot \textbf{m}_+$) with the total matter displacement $\bPsi_m$, while the right panel shows contractions with the total matter bias $b_1 \delta_m$.
    }
    \label{fig:compton_pk}
\end{figure}

A final caveat occurs when the non-gravitational forces on the tracer, $\vb{F}_{b,g}$ are nonzero. The integrated effect of such forces on $\bPsi_{r,g}$ must then be accounted for. For example, when dealing with baryons and dark matter, the effects of Compton drag on large scales are non-negligible. In this case, since the Compton drag force is proportional to the total-matter displacement, the two-point functions in Eq.~\ref{eqn:defs} involving $\vb{\Delta}^r$ will gain a contribution proportional to $\vb{\Delta}^m$ (Fig.~\ref{fig:compton_2pt}). Such contributions can be non-negligible at large scales and can dominate in the  contributions to the power spectrum proportional to $f_g$ at low wavenumber (Fig.~\ref{fig:compton_pk}). Importantly, terms proportional to $b_\pm$ are unaffected since they are related only to the primordial modes $\vb{r}_\pm$, breaking the degeneracy between $f_g$ and $b_{+}$. Since the difference between these terms is proportional to the total-matter component, this difference can alternatively be absorbed into the total-matter bias $b_m$ \cite{Schmidt17}. Comparisons of these terms with and without Compton drag are shown in Figure~\ref{fig:compton_pk}. Comparing the $f_g$ contribution with and without Compton drag we see, as expected, that renormalizing the linear total-matter bias $b_1$ to include a contribution proportional to $f_g D_{\rm CD}(z)$ (purple dotted curve) is sufficient to account for the non-gravitational Compton drag contributions.

\subsection{Cross-Spectra of different tracers and IR Resummation}
\label{sec:cross_spectra}

So far we have dealt only with tracer auto-spectra. The situation for cross-spectra is complicated by the non-cancellation of the IR-exponent at small separations, $q$. For two generic fluids, $X$ and $Y$, such that $\bPsi_{X,Y} = \bPsi_m + f_{X,Y}\bPsi_r$, the cross spectrum will take the form as in Equation~\ref{eqn:dens_eq}:
\begin{equation}
    P_{XY}(k) = \int d^3q \, e^{i \mathbf{k} \cdot \bq} e^{-\frac{1}{2}k_i k_j A^{XY}_{ij}} \Big[ \cdots  \Big],
\end{equation}
where the exponentiated two-point function $A^{XY}$ is given by
\begin{equation}
    A^{XY}_{ij}(q) = \Big( \langle \bPsi^X_i \bPsi^X_j \rangle + \langle \bPsi^Y_i \bPsi^Y_j \rangle - 2 \langle \bPsi^X_i \bPsi^Y_j \rangle \Big) + \Big( 2\langle \bPsi^X_i \bPsi^Y_j \rangle - 2\langle \bPsi^X_i(\vb{q}) \bPsi^Y_j(0) \rangle \Big),
    \label{eqn:ir_nocancel}
\end{equation}
and expectation values of point operators are displayed without arguments. Both terms in parentheses on the RHS of Equation~\ref{eqn:ir_nocancel} are well-defined and invariant under generalized Galilean transformations; however the second term vanishes as $q \rightarrow 0$ while the first does not\footnote{A similar non-cancellation occurs in the modeling of BAO reconstruction, where the cross-term between the `displaced' and `shifted' fields exhibits the same behavior \cite{PWC09,Whi15}.}. As first noted in Ref.~\cite{Lewandowski15}, this is in contrast to the single-fluid case where $A_{ij}$ had to vanish at small scales due to Galilean invariance. 

In principle, the non-cancellation discussed above will introduce a large scale damping in the power spectrum at scales proportional to the difference $|\bPsi^X -\bPsi^Y|^2$. However, since $\bPsi^{X,Y}$ are both expected to have the same coefficient in the total-matter component (i.e.\ unity) this difference squared will generically be proportional to $(f_X -f_Y)^2 \mathcal{O}(\bPsi_r^2)$, and thus is suppressed by about four orders of magnitude relative to the Zeldovich displacement, $\Sigma^2$, at the redshifts with which we are concerned ($z < 10$). On the other hand, while differential streaming is expected to damp cross spectra negligibly even if $f_X$ is of order unity, as discussed in the previous section it will still generate an observable effect degenerate with the relative bias $b_+$.

\subsection{Higher Order Bias}
\label{ssec:higher_order_bias}

Thus far we have not discussed the fact that any perturbative model should be considered an effective field theory, working up to some scale $\Lambda$ \cite{BNSZ12,CHS12,Sen14}. This forces us to introduce a set of counterterms that remove the small scale sensitivities of the perturbative calculations. For instance all the $A_{ij}(q)$ terms contain a zero-lag piece computed at zero separation, \ie $q=0$, where perturbation theory breaks down. In the single fluid case, this UV-sensitivity is renormalized to lowest order in the power spectrum by a counterterm $c_s k^2 P_{ZA}(k)$ \cite{PorSenZal14,VlaWhiAvi15}, where the free parameter $c_s$ has to be matched to simulations or data. The same structure of the counterterms appears in the two fluid scenario: for instance, the $A^{ab}_{ij}(q)$ required to calculate auto and cross spectra feature the same UV-sensitive contributions as $q \rightarrow 0$, requiring one value of $c^a_s$ for each species. In principle, terms in the equations of motion due to the relative component will add additional UV sensitivities to our predictions; in practice, however, such contributions are subdominant in the dynamics of the relative component and negligible for the total-matter component (Appendix~\ref{sec:higher_order_eom}). To the extent that these contributions can be ignored, then, the two-fluid equations of motion can be renormalized identically to the single fluid case with one set of counterterms for each species or tracer. As counterterms have minor impact on BAO scales, and are anyway fitted to the data in both the single and multiple fluid cases, we do not include them in the Fisher calculation in the next section.

We have equally refrained from discussing bias beyond linear order. As in the equations of motion, contributions beyond first order in the linear power spectrum proportional only to the total-matter component can be added consistently as in the single-fluid case, and we will ignore small nonlinear contributions proportional to one or more powers of the relative component\footnote{A proper accounting of such terms would in addition require solving the relative-component equations of motion to beyond linear order, which is beyond our present scope.}. However, one exception must be made: operators involving the relative-velocity between the baryon and dark matter squared, which, despite being at second order in the relative component, can be non-negligible due to their distinct dimensional scaling \cite{Dalal10,Yoo11,Blazek16, Schmidt16}. Such contributions were the focus of the first studies of bias \cite{Dalal10,Yoo11,Blazek16} in the two-fluid picture, and we will show how their calculation fits naturally into the Lagrangian framework. For a discussion of other second order bias operators see Appendix \ref{sec:second_order}.

At second order in the bias expansion we can write
\begin{align}
    F_g[\bPsi_m,\,\bPsi_r|\,\vb{q}] \supset b_{v}\sigma_{v_r}^2\ \frac{[\vb{v}_b(\vb{q})-\vb{v}_c(\vb{q})]^2}{\sigma_{v_r}^2} = b_{v}\sigma_{v_r}^2 \frac{\vb{r}_-(\vb{q})^2}{\sigma_{r_-}^2}
\equiv c_- [\vb{r}_-(\vb{q})]^2 \end{align}
where $\sigma_{v_r}^2$ is the 1-point variance of the relative velocities and $\sigma_{r_-}^2 = (1+z)^{-2} H_0^{-2} \sigma_{v_r}^2$.
As several authors \cite{Dalal10,Schmidt16} have pointed out, baryon-dark matter relative velocities can be quite large at the time when the first halos and galaxies form, which could result in a large value of $b_v$ for their late time descendants. The value of $b_{v} \sigma_{v_r}^2$ can be as large as 0.01, which will make this contribution at second order in the power spectrum larger than the $b_-$ terms, even on linear scales. It is however worth remembering that a value of $b_{v} \sigma_{v_r}^2\simeq 10^{-5}$ is also plausible, which would substantially reduce the importance of this contribution.

To consistently compute the power spectrum contributions due to $c_- \sim b_{v^2}$ we must go beyond the Zeldovich approximation. Up to 1-loop in Lagrangian perturbation theory we have to compute 4 new terms to properly include the new bias parameter $c_-$. Beyond these, terms proportional to $c_-^2$ can be safely neglected as they are $\mathcal{O}(P_{\nabla \vb{r}_-^2}^2)$. For the same reason we drop all the terms proportional to $b_{\pm}c_-$, as well as contributions of the relative component to the equations of motion. This leaves us with contributions proportional to $c_-$, $b_1 c_-$, $b_2 c_-$, and $b_{s^2} c_-$.

The first of these, proportional to $c_-$, contains a 1-loop contribution and is given by 
\begin{align}
P_{gg}(k)\supset c_- \int \dq \ZA \, \Big( 2 i k_i \mathcal{U}_i(q) - \frac{1}{2} k_i k_j A^{m-}_{ik} A^{m-}_{jk} \Big),
\end{align}
where we have defined $A^{m-}_{ij} = \avg{\Delta_{m,i}\ (\vb{r}_{-,2}-\vb{r}_{-,1})_j}$ and the 1-loop contribution from the second-order Lagrangian displacement $\bPsi^{(2)}$ enters as
\begin{align}
     \mathcal{U}_i(q) \equiv \avg{\Delta^{(2)} \vb{r}_{-,1}^2} = \hat{q}_i \int \frac{\mathrm{d}k}{2 \pi^2}\, k^2 Q_{v^2}(k) j_1(kq) \,
\end{align}
The kernel $Q_{v^2}$ is derived in Appendix~\ref{sec:calc_details}.

The remaining terms do not contain loop contributions and follow straightforwardly from evaluating the second and third cumulants in \eq{eqn:power_spec} within the Zeldovich approximation. These are those proportional to the first order bias:
\begin{align}
    P_{gg}(k)\supset  2 i k_i\ b_1 c_- \int \dq \ZA \, A^{m-}_{ij}(q)\ U^{-m}_j(q)\,,
\end{align}
second order bias:
\begin{align}
P_{gg}(k)\supset 2 b_2 c_- \int \dq \ZA U^{m-}_i(q)\ U^{m-}_i(q),
\end{align}
and shear
\begin{equation}
    P_{gg}(k)\supset  4 b_{s^2} c_- \int \dq \ZA W^{s-}_{ijk}(q)\ W^{s-}_{ijk}(q),
\end{equation}
where we have defined the 2-point functions $U^{-m} \equiv \avg{\vb{r}_-(q) \delta_m(0)} = U^{m-}$ and $W^{s-}_{ijk}(q) = \avg{s_{ij}(q) \vb{r}_{-,k}(0)}$. Details of the above calculation can be found in Appendix \ref{sec:calc_details}.

\begin{figure}
    \centering
    \includegraphics[width=\textwidth]{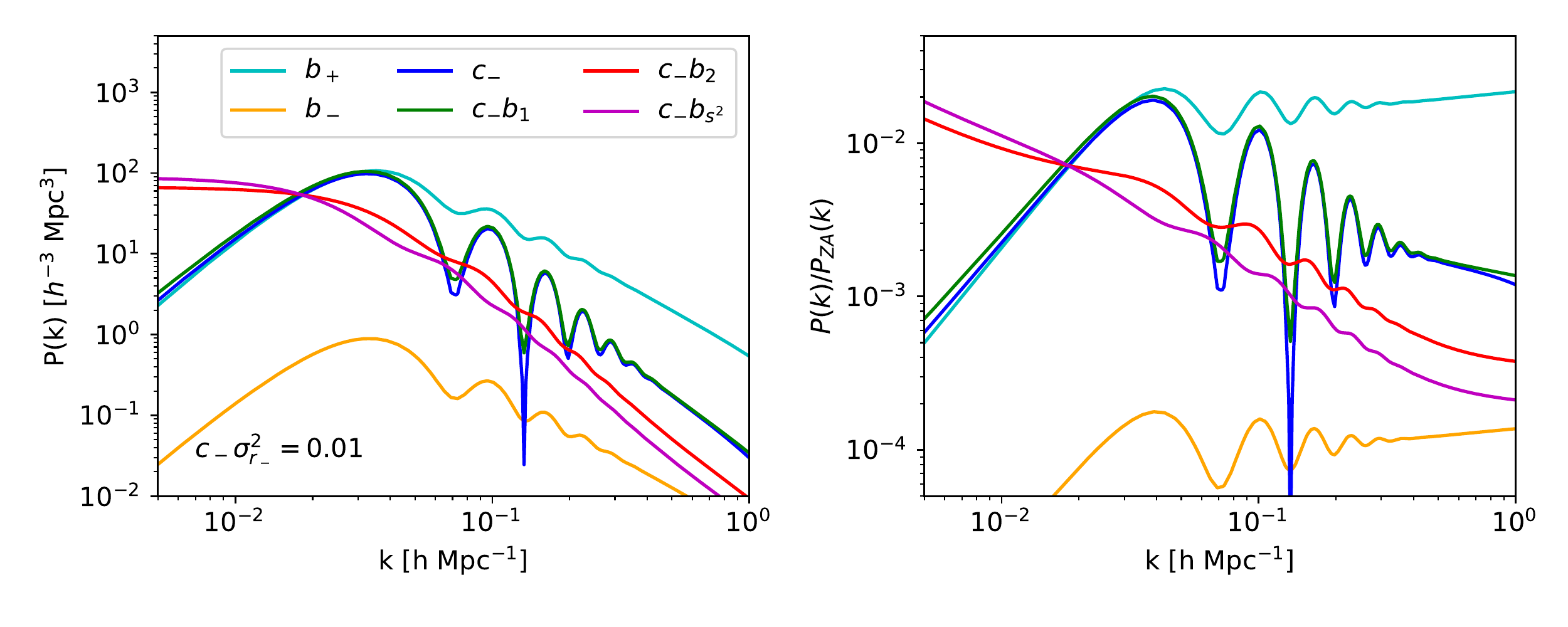}
    \caption{Contributions to the Zeldovich galaxy power spectrum from relative velocity bias at second order. All biases are set to unity except for $c_-$, which is set such that $b_v \sigma_{v_r}^2 = 0.01$---in this case, the contributions from $b_{v^2}$ are seen to be quite comparable to those from $b_+$, and moreover exhibit BAO ``wiggles'' far more prominently than does the regular ZA contribution.}
    \label{fig:rel_vel_bias}
\end{figure}

The contributions proportional to $c_-$ and their comparison with the 1-piece in \eq{eqn:pk_zeldovich} and with the $b_\pm$ ones computed in the previous section is shown in Fig~\ref{fig:rel_vel_bias} for $z=1.2$, assuming $b_v \sigma_{v_r^2}^2 = 0.01$.
The $c_-$ terms are indeed larger than the $b_-$ terms on most scales, but still subdominant compared to the $b_+$ terms. Notably, the $c_-$ terms feature significantly larger oscillatory features than contributions from $b_\pm$, with minima that differ from maxima by more than an order of magnitude. 

\section{Degeneracies and bias to BAO}
\label{sec:fisher}

\begin{figure}
    \centering
    \includegraphics[width=\textwidth]{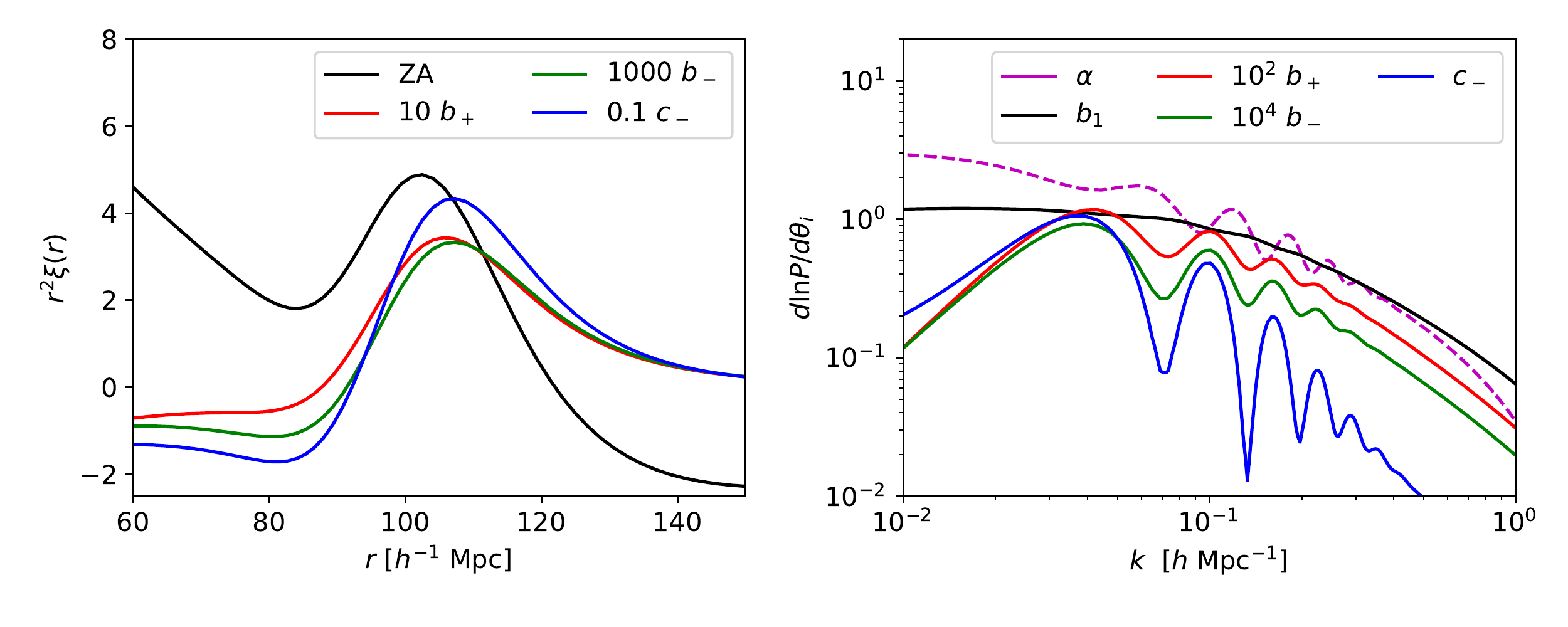}
    \caption{(Left) Contributions to the $z = 1.2$ correlation frunction from the various relative component biases, multiplied by constant factors for ease of comparison. All contributions have prominent features at the BAO scale, reflecting their origin in early-universe acoustic oscillations. (Right) Derivatives of the power spectrum with respect to these parameters and the BAO scale parameter $\alpha$ at $z = 1.2$, with $b_m = 0.5$, $b_2 = 0.2$, $b_+ = 1$, $b_- = 7$ and $c_- \sigma_{r_-}^2 = 0.01$. Despite the fact that all these templates feature prominent oscillations, they nonetheless possess distinct scale dependence. Note that some of the derivatives have been multiplied by powers of ten for ease of comparison.
    }
    \label{fig:fisher_derivs}
\end{figure}

Baryon acoustic oscillations (BAO) in the photon-baryon fluid before combination imprint a characteristic clustering scale in the distribution of galaxies that can be used as a standard ruler to constrain the cosmic expansion history \cite{PDG18}.  In general this method is regarded as highly robust as it probes very large scales which are largely unaffected by astrophysical processes.  However, relative component contributions to the two-point function also occur on very large scales and their oscillatory features, although arising from the same physical process of the standard BAO features in the matter density power spectrum, could bias our estimates of the distance scale if not properly taken into account \cite{Blazek16,Beutler2017,Slepian}. Indeed, as shown in the left panel of Fig.~\ref{fig:fisher_derivs}, all the relative component contributions we have considered show distinct features around the BAO peak.

The extent to which contributions from the relative component can contaminate measurements of the BAO scale can be estimated using the Fisher matrix formalism \cite{Tegmark97}. The galaxy overdensity has a covariance that is diagonal in Fourier space and given by the power spectrum plus shot noise, $\hat{P}_{gg} =P_{gg}(k)+\bar{n}^{-1}$; for the parameters $\{\theta_i\}$, the Fisher matrix is given by
\begin{equation}
    F_{ij} = V_{obs} \int \frac{d^3 k}{(2 \pi)^3} \, \frac{1}{2} \frac{\partial \ln \hat{P}_{gg}(k)}{\partial \theta_i} \frac{\partial \ln \hat{P}_{gg}(k)}{\partial \theta_j},
    \label{eq:Fisher}
\end{equation}
where $V_{obs}$ is the observed volume. For simplicity we neglect redshift space distortions and focus only on the isotropic BAO signal, though we will comment on how our Lagrangian analysis can be naturally extended to redshift space in the final paragraph. We model the power spectrum using the two-fluid Zeldovich terms derived above and include matter contributions up to one loop (see e.g. \cite{VlaCasWhi16}), including contributions from the quadratic Lagrangian bias $b_2$. We consider only scales between $k_{\rm min}=10^{-2}\,\kMpc$ and $k_{\rm max} = 0.25\,\kMpc$, and fiducial value of $b_1=0.53$ and $b_2=0.2$. The number density of galaxies is $\bar{n} = 4.2\times10^{-4}\ \icMpc$ and we assume $V = 5 \cGpc$.
These numbers are chosen to be similar to what galaxy surveys like DESI \cite{DESI} or Euclid \cite{EUCLID18} are expected to measure, and in particular are based off the expected DESI ELG population at $z = 1.25$ in a bin of width $\Delta z = 0.1$ and 14,000 square degrees of observation.

To quantify the potential impact of the relative component on standard BAO analyses, we will compare two models of the power spectrum within the Fisher formalism: the ``correct'' model $\textbf{M}_1$, which is a function of all total-matter and relative component biases, and the nested ``standard'' model $\textbf{M}_0$, wherein the relative component biases are set to zero (i.e.\ $b_\pm,\ c_- = 0$). The observed power spectrum is in addition a function of the BAO scaling parameter $\alpha$ such that
\begin{equation}
    P_{gg}(k,z,\alpha, \textbf{M}) = \alpha^{-3} P_{gg}\left(\frac{k}{\alpha},z,\textbf{M}\right).
\end{equation}
The derivative of the baseline galaxy power spectrum with respect to the parameters is shown in \fig{fig:fisher_derivs}. These templates all show oscillatory features of roughly the same frequency as the BAO scale but exhibit distinguishable scale dependence.
For reference, applying \eq{eq:Fisher} returns sub-\% error on the BAO scale, with $\sigma_\alpha = 0.9\%$, for the standard analysis using $\textbf{M}_0$.

\begin{figure}
    \centering
    \includegraphics[width=\textwidth]{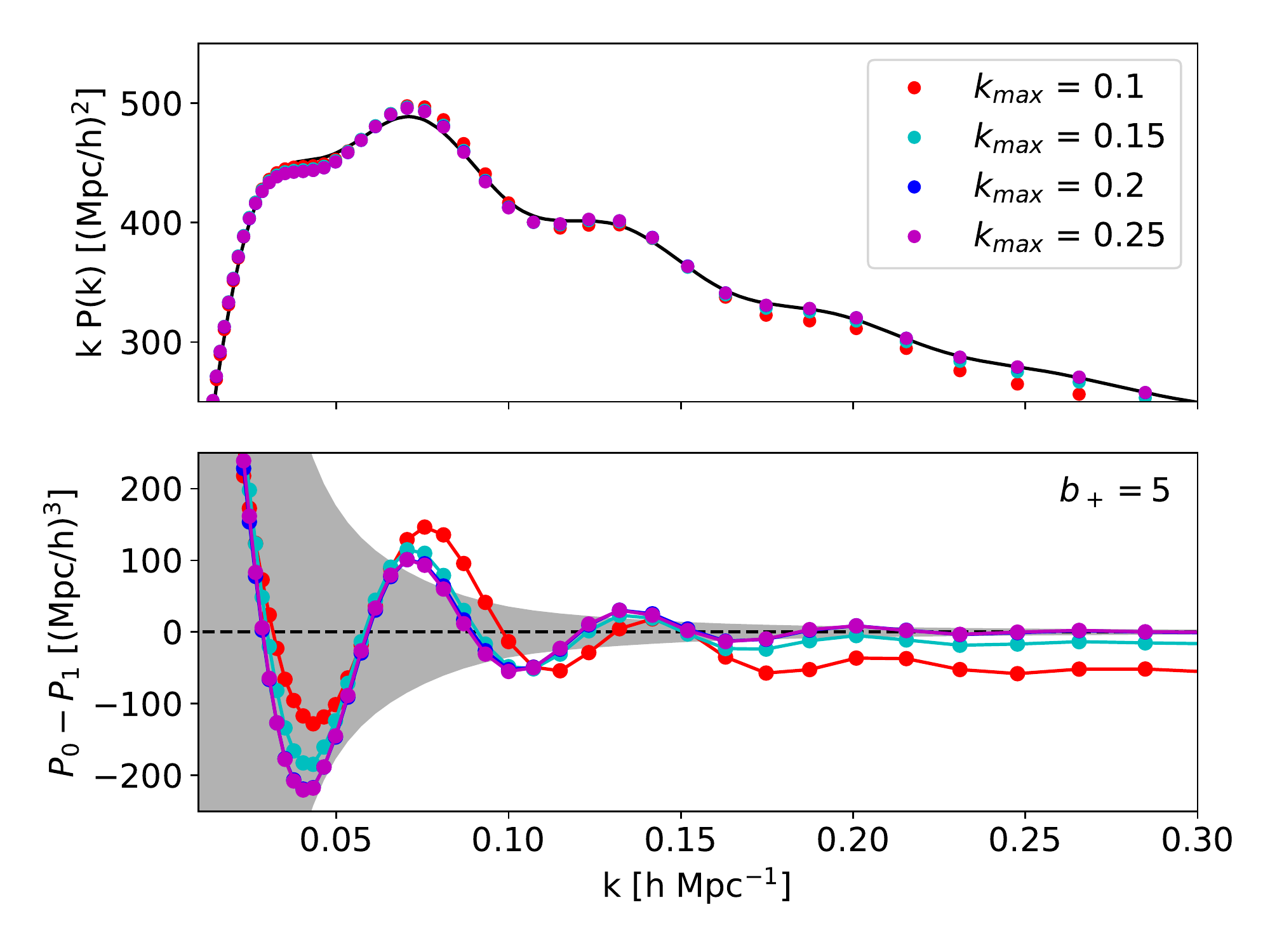}
    \caption{(Top) Best fit power spectra using the total-matter-component-only model, $\textbf{M}_0$, for a universe where $b_+ = 5$ with varying maximum fitted wave numbers $k_{\rm max}$. (Bottom) Residuals of the above fits, compared to expected errors ($\Delta \ln k = 0.06)$, shaded in gray. Fitting over too narrow a range ($k_{\rm max} = 0.1 \kMpc$) results in a highly biased phase, while fits using larger wave number ranges covering more than one BAO wiggle are essentially in phase. The remaining oscillating residuals significantly exceed the expected error and are due to lack-of-fit for the oscillations in the relative component.}
    
    \label{fig:shifted_fit}
\end{figure}

We can now compute the systematic shifts in $\alpha$ that would be incurred by neglecting the relative component, i.e. by fitting to $\textbf{M}_0$. For convenience, we will split the parameters in $\textbf{M}_1$ into $\theta = (\phi_a, \psi_\sigma),$ where $\phi_a$ with Latin indices are the BAO scale and total-matter parameters and $\psi_\sigma$ with Greek indices are the relative component biases, such that $\textbf{M}_0$ is given by $\theta = (\phi_a, \psi_\sigma=0).$ In this language the shift in $\alpha$ and $b_1$ due to using the standard model can be calculated to first order as \cite{Taylor07}
\begin{equation}
    \delta \theta_a = - (F_0)^{-1}_{ ab} G_{b \sigma} \delta \psi_\sigma, \qquad a, b = \alpha, b_1, \; \sigma = b_{\pm}, c_-.
    \label{eqn:shift_eqn}
\end{equation}
Here $F_0$ and $G$ are respectively diagonal and off-diagonal blocks of the full Fisher matrix $F = F(\theta_0)$ calculated at the best fit parameters $\theta_0$ for the full model $\textbf{M}_1$, such that $F_{0,ab} = F_{ab}$ and $G_{b\sigma} = F_{b\sigma}$, and $\delta \psi$ is the deviation of $\psi$ in the standard cold dark matter only model $\textbf{M}_0$ from $\vb{M}_1$, i.e. $\delta \psi = - \psi_0.$

\begin{figure}
    \centering
    \includegraphics[width=\textwidth]{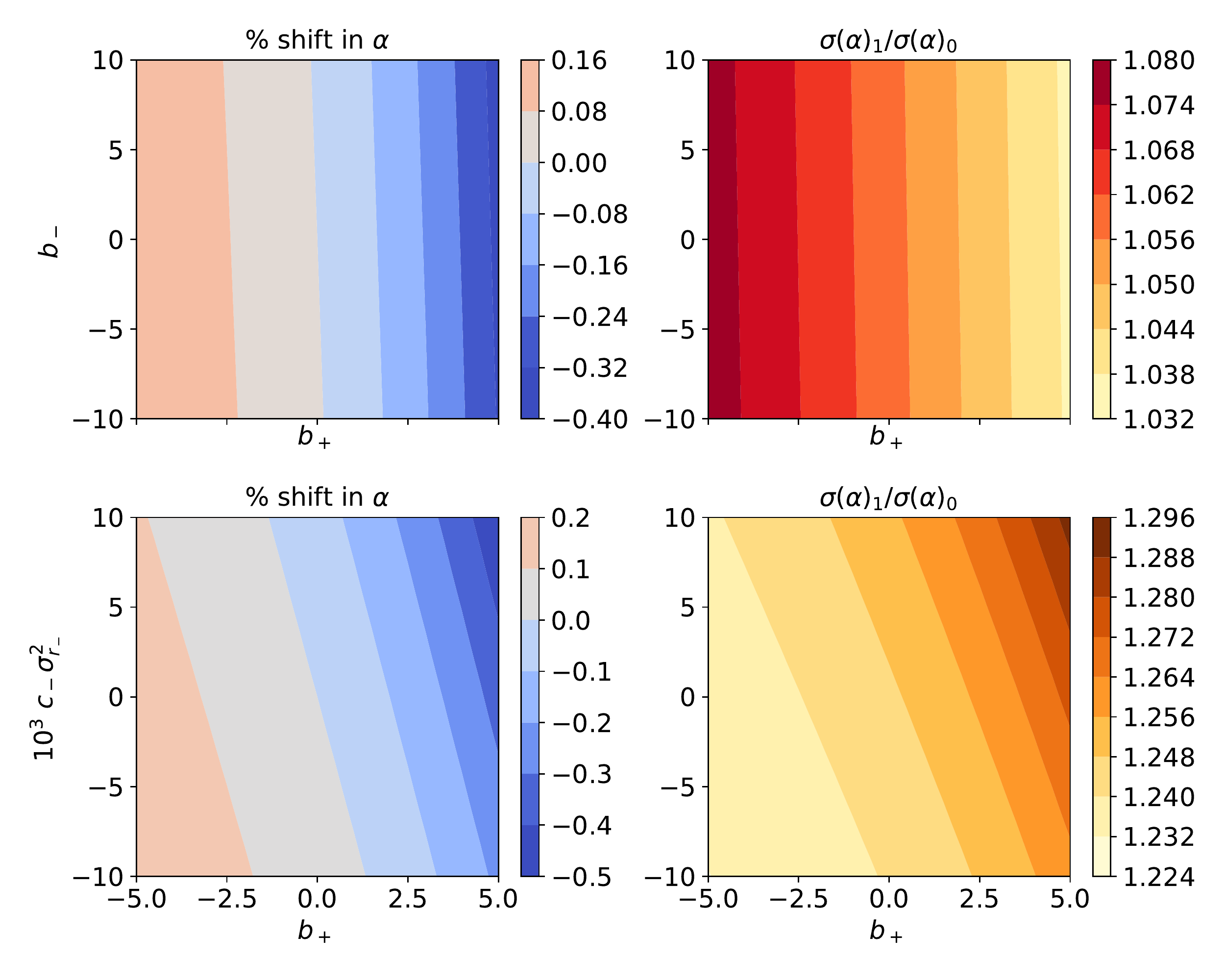}
    \caption{(Top Left) Shift in measured $\alpha$ when neglecting relative component biases as a function $b_\pm$ in the absence of $c_-$. While $b_-$ contributes negligibly, $b_+ = 5$ produces a shift up to a $0.4\%$. (Top Right) Ratio of error bars in $\alpha$ when marginalizing over $b_\pm$ vs. when they are kept fixed at zero, such that the best-fit value of $\alpha$ is biased in the latter case. In the latter case the forecast takes into account the shift away from the true value due to incorrect model assumptions.
    (Bottom Row) Same as the above, but with $c_-$ added as a nonzero parameter in $\textbf{M}_1$. We have set the true $b_- = 0$ for convenience but marginalize over it to calculate uncertainties. While even $c_- \sigma_{r_-}^2 = 0.01$ contributes only a tenth of a percent to the shift in $\alpha$, the error bars are inflated relative to the top row by up to twenty percent. We assume $k_{\rm max} = 0.25 \kMpc$ throughout.}
    \label{fig:fisher_pm}
\end{figure}

As a simple first example, we consider a toy-model Universe in which the only relative contribution is $b_+$. Figure~\ref{fig:shifted_fit} compares the ``true'' power spectrum, $P_1(k)$, assuming $b_+ = 5$, with best fits to the power spectrum in a dark matter only universe $P_0(k)$, described by the model parameters $\textbf{M}_0$, where the values of $\alpha$, $b_1$, $b_2$ are shifted from their true values according to Equation~\ref{eqn:shift_eqn}. Different values of the maximum wave number $k_{\rm max}$ included in the Fisher calculation are shown with different lines. For $k_{\rm max} = 0.1 \kMpc$, we find a significant departure in phase between the two models, compared to higher limiting wavenumbers, as evident from the phase of the residual in the bottom panel. Beyond $k_{\rm max} = 0.15 \kMpc$ there are sufficient BAO wiggles that the phase of the residuals are essentially locked. We caution that the same exercise repeated with both matter and relative terms in the Zeldovich approximation can lead to wide swings in the BAO scale $\delta \alpha$ as a function of $k_{\rm max}$. This can be understood as follows: at $k \gtrsim 0.1 \kMpc$, $b_+$ contributes both oscillatory behavior and a broadband shape identical to the total matter component. The latter is essentially an amplitude change and can be roughly cancelled by a shift $\delta b_m,$ which it is thus fixed independently of $k_{\rm max}$. This then requires $\delta \alpha$ to shift with $k_{\rm max}$ as more oscillations are included until the oscillations in $\vb{r}_+$ relative to $\vb{m}_+$ are damped at large $k$ (Figure~\ref{fig:tf_ratios}). This broadband effect is ameliorated by including nonlinear terms for BAO measurements, but the partial degeneracy of $b_+$ with the power spectrum amplitude likely implies that ignoring two-fluid effects may affect measurement of the amplitude of the power spectrum (though this effect will also be partially mitigated by redshift-space distortions).

\begin{figure}
    \centering
    \includegraphics[width=\textwidth]{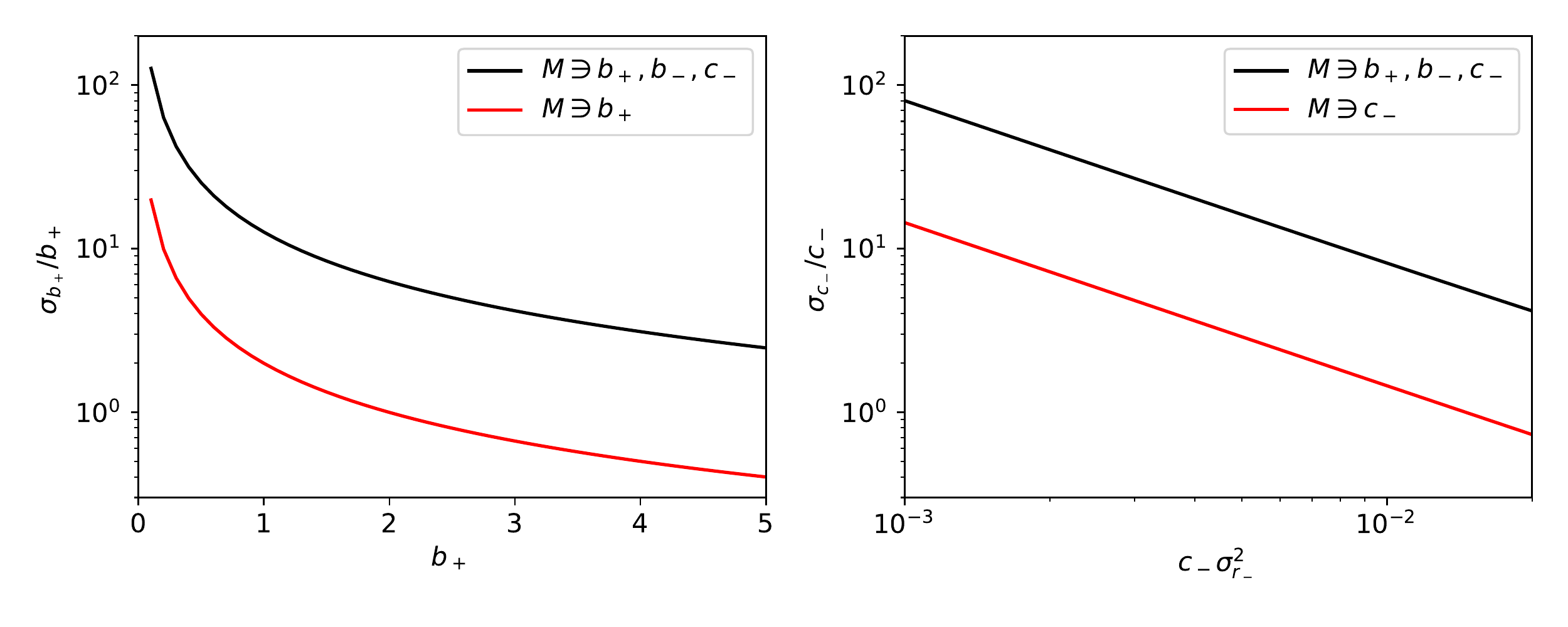}
    \caption{Constraints on $b_+$ and $c_-$ in our fiducial setup if only each respective parameter can be varied (black), and if all relative parameters are simultaneously marginalized over (red). Notably, when the full model is taken into account detecting the relative velocity effect ($c_-$) will require up to ten times more signal to noise.}
    \label{fig:detectability}
\end{figure}

The same formalism can be applied to more realistic bias models. In the upper left panel of Figure~\ref{fig:fisher_pm} we consider the case when the observed power spectrum contains  nonzero values $b_\pm$ and $c_- = 0,$ and forecast the shifts in $\alpha$ due to the wrong assumption of $b_\pm=0$. Due to the small size of the $b_-$ contributions (see Figure \ref{fig:Pks}), we expect shifts in BAO inferred distances to be dominated by $b_+$, and this is indeed what we find, contours of constant $\delta \alpha$ are almost independent of $b_-$ even when $|b_-| = 10.$  On the other hand, we see that values of $b_+ \sim 5$ shift the measured $\alpha$ by up to $0.4\%$, close to half of the error on $\alpha$ expected when using $\textbf{M}_0$.

However, the physics behind the relative components is quite well understood and can be easily included in Fisher forecasts or power spectrum analyses. Indeed, as seen in Figure~\ref{fig:fisher_derivs}, the templates for the various relative biases and $\alpha$ have distinct shape and could be distinguishable depending on the noise level of the measurements. The upper right plot in Figure~\ref{fig:fisher_pm} shows the increase in $\sigma_\alpha$ induced by marginalizing over $b_\pm$ in universes where $b_\pm$ and $b_v$ are not necessarily nonzero \footnote{The nonzero $b_{\pm,v}$ produce shifts in the measured $\alpha, b_m$ when using $\textbf{M}_0$, which must be taken account when computing $\sigma_\alpha.$ To first order, the shifted Fisher matrix is given by $F_0$.}. The total loss of constraining power is modest, with less than 10\% worse error bars even after marginalizing over two extra parameters.
In both the computations of the shifts in $\alpha$ and the increase of $\sigma(\alpha)$, the volume of the survey does not enter, and the final results depend only on the shot noise levels.

In the lower set of plots in Figure~\ref{fig:fisher_derivs}, we repeat the same exercise described above including $c_-$ as an extra free parameter. Since $b_-$ is  irrelevant for the final results we set it to zero (but still marginalized over it).
We find that $b_+$ and $c_-$ are anti-correlated, with larger shifts compared to the $b_\pm$ case, but $\delta \alpha/\alpha \le 0.5\%$ in all cases. Marginalizing over the extra parameter $c_-$ results in a 20-30\% increase in $\sigma(\alpha)$, which is still benign for BAO constraints.
Our results therefore advocate for the implementation of relative component biases, at least of $b_+$ and $c_-$, in standard BAO data analysis of the galaxy power spectrum or correlation function.

Finally, in Figure~\ref{fig:detectability} we investigate the detectability of the two-fluid effects in the same setup. On their own, both $b_+$ and $c_-$ become 1$\sigma$ detectable at the upper end of our explored parameter ranges, shown as the red lines in Figure~\ref{fig:detectability}. However, once all three relative bias parameters are marginalized over, the black set of curves in Figure~\ref{fig:detectability}, neither will be detectable within our fiducial volumes, with $c_-$ in particular at $0.1 \sigma$, well out of reach even if all the DESI redshift bins are combined.

\section{Conclusions}
\label{sec:conclusions}

The large scale structure of the universe, whose formation is dominated by the dynamics of gravitational collapse, is one of the premier probes into fundamental physics. At subleading order, the presence of multiple particle species, broadly categorized into cold dark matter, baryons and neutrinos, with distinct properties beyond their shared gravitational attraction, can present additional features in this structure, which will become increasingly important as future surveys push to higher precision. In particular, relative perturbations between baryons and cold dark matter are prominent at the same scale as baryon acoustic oscillations and have the potential to cause systematic biases in future BAO measurements.

In this paper, we develop the Lagrangian formalism to calculate the clustering of biased tracers in the presence of multiple fluids, focusing specifically on the two-fluid scenario with dark matter and baryons. The Eulerian description of two-fluid dynamics has been studied extensively in the past and we make contact with previous work as appropriate throughout the text. LPT includes an automatic resummation over long-wavelength bulk flows and is thus able to accurately capture the shape of BAO features for biased tracers. In addition, LPT naturally maps bias terms from their initial Lagrangian positions to advected Eulerian positions, in contrast to Eulerian theory in which advective terms must be put in by hand, thereby simplifying the treatment of bias as responses to linear initial perturbations.

The presence of two fluids introduces terms beyond those encountered in traditional single fluid cosmological perturbation theory, with modifications in both the bias expansion and tracer advection. In the former, the generalized Galilean invariance that restricted the bias to contain only second derivatives of the gravitational potential in the single fluid case, allows terms including relative overdensities and velocities between different species.  In the latter, initial relative displacements between various species are preserved under free fall and present an additional source of bias. Large scale non-gravitational forces such as Compton drag induced by the CMB can introduce additional corrections. We formulate modifications to tracer bias and advection in terms of three initial modes, constants of motion in the linear equations of motion, which roughly correspond to the initial total-matter displacement field and the relative displacement and velocity fields between dark matter and baryons.

We explicitly calculate the galaxy auto-power spectrum in the Zeldovich approximation within this formalism. Cross correlations between the relative modes introduce eight terms linear in the power spectrum---however, those quadratic in the relative component are suppressed by four orders of magnitude relative to the single fluid terms at low redshifts relevant for the next generation of galaxy surveys. Comparing to the Eulerian result explicitly to first order in the power spectrum, we find that the Eulerian relative component bias corresponds to linear mixtures of the Lagrangian bias, with modifications to the tracer advection entering both the Eulerian relative overdensity bias and the Eulerian relative velocity divergence bias. We then take up the calculation of cross spectra, finding a large scale damping due to an IR noncancellation in the relative component that is nonetheless negligibly small on perturbative scales. We also briefly discuss higher order corrections to the equations of motion in the presence of two fluids from an effective theory point of view, and perform an example one loop calculation for the relative velocity effect ($\propto \vb{v}_r^2$).

We conduct an exploratory analysis into whether two-fluid effects can cause systematic biases in measurements of the BAO scale. Taking the example of DESI ELGs at $z = 1.25,$ we show that while ignoring two-fluid effects can lead to systematic shifts in the measured BAO scale as large as half a sigma, properly marginalizing over these effects induces less than ten percent loss in precision for a wide range of bias values. Since the scale dependence of the underlying physics is well understood, these results advocate for including two-fluid terms at linear order in future analyses. The dominant relative bias term ($\propto b_+$) does not fall quadratically with the growth factor like the total-matter contributions, and we therefore expect the relative bias signal as a fraction of total power to scale with redshift as $D^{-1}_{+}(z)$ and become proportionally more significant for surveys (such as the proposed Stage II 21-cm survey \cite{CVDE-21cm}) at higher redshifts. Studies of more highly biased tracers such as DESI quasars \cite{DESI}, for which the total-matter contributions are correspondingly larger, will on the other hand be less influenced by the relative bias for similar reasons.

While the Lagrangian picture is a natural playground for their study, in this paper we have opted not to study redshift space distortions (RSD). We note, however, that of the two relative components, $\vb{r}_+$ is dominant but stationary while $\vb{r}_{-}$ is so small as to be essentially negligible--- two-fluid impacts should thus have a relatively small impact on RSD. However, as noted in the previous section, since the dominant relative component contribution $b_+$ is somewhat degenerate with the overall power spectrum amplitude, it is possible that two-fluid effects could hinder the accuracy of $f \sigma_8$ measurements beyond the percent level. We will return to this issue in future work.

\section*{Acknowledgments}
We thank Jonathan Blazek and Jahmour Givans for helpful comments. SC is supported by the National Science Foundation Graduate Research Fellowship (Grant No. DGE 1106400) and by the UC Berkeley Theoretical Astrophysics Center Astronomy and Astrophysics Graduate Fellowship.
M.W.~is supported by the U.S.~Department of Energy and by NSF grant number 1713791.
This work made extensive use of the NASA Astrophysics Data System and of the {\tt astro-ph} preprint archive at {\tt arXiv.org}.

\appendix

\section{Redshift dependence and size of the of bias parameters}
\label{sec:running_bias}

The bias expansion in Equation~\ref{eqn:bias_exp} has an implicit dependence on the initial redshift $z_i$ that must be taken into account to reach consistent conclusions. Since the initial conditions mix at most linearly, no information can be lost by choosing one initial time $\tau_i$ over another; for example, the sensitivity of halos to the relative velocity divergence after reionization, which contains a contribution from the total matter overdensity (Eq.~\ref{eqn:compton_coeff}), can be directly accounted for by calibrating the bias parameter for $\delta_m$ at an earlier redshift.

As a simple example we consider the redshift dependence of the relative components in the sourceless ($F_b = 0$) case. If we set our initial time at $\tau_i'$ instead of $\tau_i$ we will get
\begin{equation}
    \bPsi_r(\tau) = \Big(-\mathbf{r}_{+} + \mathbf{r}_{-} D_{r}(\tau_i',\tau_i)\Big)+ \mathbf{r}_{-} D_{r}(\tau,\tau_i') \equiv -\mathbf{r}_{+}' + \mathbf{r}_{-}' D_{r}(\tau,\tau_i').
\end{equation}
Re-expanding $F_g$ at $\tau_i'$ thus yields
\begin{equation}
    F_g(\bq) = b_1 \delta_m + b_{+}' \nabla \cdot \Big(r_{+} - r_{-} D_{r}(\tau_i',\tau_i)\Big) + b_{-}' \nabla \cdot r'_{-} + ...
\end{equation}
Since $b'$ and $b$ apply to the same field configurations at different times, they must yield the same initial overdensity $F_g$ --- this requirement can be satisfied by enforcing the differential equations
\begin{equation}
    \frac{db_{+}}{d\tau} = 0, \qquad
    \frac{db_{-}}{d\tau} = \frac{b_{+}}{a(\tau)}
    \qquad .
    \label{eqn:bias_running}
\end{equation}
Intriguingly, the presence of a relative overdensity bias can ``generate'' a relative velocity bias at later times. This can be understood as follows: the relative overdensity at late times is a linear combination of the relative overdensity and velocities at earlier times. Similar, though more complicated, versions of this relation hold when $F_b \propto \vb{m}_+$, in which case mixing of all three initial fields must be taken into account.

\section{Beyond Linear Order}
\label{sec:higher_order_eom}

\subsection{Equations of motions}

In this appendix we derive the equations of motion beyond linear order in the two-fluid scenario, and show that the nonlinear contribution of $\vb{r}_\pm$ to the total-matter component are quadratically suppressed, and that the nonlinear relative component is always sourced by at least one component of $\vb{r}_\pm$.

The Lagrangian equations of motion at higher order can be found by taking the real-space divergence of both equations in \ref{eqn:species_eom}. To do so we make use of the identities
\begin{align}
    1 + \delta^a(x,t) &= \Big\rvert\frac{\partial x^a}{\partial q}\Big\lvert^{-1}  = \mathcal{J}^a(q,t)^{-1} \\
    \nabla_{x^a} \cdot V &= \Big[\frac{\partial x^a}{\partial q} \Big]^{-1}_{ij} \frac{\partial V_i}{\partial q_j} = \big[ \delta_{ij} + \bPsi^a_{i,j}\big]^{-1} V_{i,j},
\end{align}
where the negative powers in the second line denote matrix inverses, to account for the coordinate transformations between Lagrangian coordinates $q$ and the fluid trajectories for each species $x^a = q + \bPsi^a,$ as well as the standard matrix identities $(I+A)^{-1} = I - A + A^2 - A^3 + \mathcal{O}(A^4)$ and $\det(I+A) = 1 + \text{Tr}[A] + \frac{1}{2}(\text{Tr}[A]^2-\text{Tr}[A^2]) + \mathcal{O}(A^3)$. We will neglect the effects of Compton drag, which affects the relative displacement at a few percent level even at late times and on linear scales and thus enter into our final power spectra at the same order of magnitude as the relative component squared, and assume potential flow.

The above equations imply that the Lagrangian equations of motion for the fluid displacements of a species $a$ at n$^{th}$ order takes the generic form
\begin{equation}
    D \bPsi^{(a,n)}_{i,i} = - \sum_{m=1}^{n-1} F^{(a,n-m)} D \bPsi^{(a,m)} + \frac{3}{2} \mathcal{H}^2 \Omega_m \sum_{a'} w_{a'} \Big(\frac{\mathcal{J}-1}{\mathcal{J}}\Big)^{(a',n)},
\end{equation}
where the superscript $(a,n)$ denotes the species and order of each term, the derivative operator $D$ is defined such that $DX = X''+\mathcal{H}X'$, and the $F^{(a,n)}$'s are kernels composed of displacements of the species $a$ at order $n$ and below. Switching to the total matter and relative components, we have
\begin{align}
     D \bPsi^{(m,n)}_{i,i} &= - \sum_{a'} w_{a'} \sum_{m=1}^{n-1} F^{(a',n-m)} D \bPsi^{(a',m)} + \frac{3}{2} \mathcal{H}^2 \Omega_m \sum_{a'} w_{a'} \Big(\frac{\mathcal{J}-1}{\mathcal{J}}\Big)^{(a',n)} \label{eqn:m_highorder} \\
      D \bPsi^{(r,n)}_{i,i} &= - \Bigg(\sum_{m=1}^{n-1} F^{(c,n-m)} D \bPsi^{(c,m)} - \sum_{m=1}^{n-1} F^{(b,n-m)} D \bPsi^{(b,m)}\Bigg).
      \label{eqn:r_highorder}
\end{align}
We can derive some elementary properties of these equations without solving for their particular forms using symmetry arguments. Noting that the RHS of \ref{eqn:m_highorder} is symmetric under species index exchange ($b \leftrightarrow c$) while \ref{eqn:r_highorder} is antisymmetric, we can conclude that (1) the first relative contribution to the total matter EOM at each order must be of order $\mathcal{O}(\vb{r}_\pm^2)$, and all subsequent contributions suppressed by further even powers of the linear relative component and (2) the relative component is always sourced by at least one power of $\vb{r}_\pm$, since the total matter component is even under this swap while the relative component is even. Note that (1) implies that the dynamics of the total-matter displacement are affected by the relative component only at the percent-of-a-percent level, and (2) implies that the nonlinear relative displacement is never less suppressed in $\vb{r}_\pm$ than the linear solution.. A similar result occurs in Eulerian theory, as described in \cite{Lewandowski15}. 

For completeness, the explicit second and third order equations of motion for the relative component are, up to first order in the linear relative perturbation 
\begin{align}
    D\bPsi^{(r,2)}_{i,i} = & \bPsi^{(r,1)}_{j,i} D\bPsi^{(m,1)}_{i,j} & \nonumber\\
    D\bPsi^{(r,3)}_{i,i} = & \bPsi^{(r,1)}_{j,i} D\bPsi^{(m,2)}_{i,j}  +  \bPsi^{(m,1)}_{j,i} D\bPsi^{(r,2)}_{i,j} + \bPsi^{(r,2)}_{j,i} D\bPsi^{(m,1)}_{i,j} \nonumber\\
    &- \Big( \bPsi^{(r,1)}_{j,k}\bPsi^{(m,1)}_{k,i} + \bPsi^{(m,1)}_{j,k}\bPsi^{(r,1)}_{k,i} \Big) D\bPsi^{(m,1)}_{i,j}, 
    \label{eqn:relative_high_order}
\end{align}
which are in-line with the symmetry arguments outlined above. The equations of motion for the total matter component at second and third order can be similarly verified to be simply the equations of motion in the one-fluid case to this order in the relative component.

\subsection{Biasing at second order}
\label{sec:second_order}

Below we list all contributions to the bias expansion up to second order in the initial fields omitting derivative corrections:
\begin{align}
    F_g &= \, b_1 \delta_m + b_{\delta_r} \delta_r +  b_{\theta_r} \theta_r \nonumber \\
        &+ \frac{1}{2} b_2 \delta_m^2 + b_{s^2} s_{ij} s_{ij} + b_{\delta_m \delta_r} \delta_m \delta_r + b_{\delta_m \theta_r} \delta_m \theta_r + b_{v_r \partial \delta_m}  (\textbf{v}_r)_i \, \partial_i \delta_m + b_{s \, \partial v} \partial_i (\textbf{v}_r)_j s_{ij} \nonumber \\
        &+ b_{v^2_r} \textbf{v}_r^2 + ....
        \label{eqn:app_bias}
\end{align}
In the main body of this paper we consider relative bias terms up to first order in the power spectrum, since even these represent only percent level effects, with the exception of the relative velocity effect $\propto \vb{v}_r^2$, which has a distinct scaling. As noted in the text, we note that the presence of Compton drag can introduce additional terms due to loss of gauge redundancy; we refer readers to the extensive discussion in \cite{Schmidt17}.

\section{Relative Velocity Bias Terms}
\label{sec:calc_details}

In this appendix we provide details for the contributions of the relative velocity bias $b_{v^2}$ at $\mathcal{O}(P^2)$ to the galaxy power spectrum. These contributions require the calculation of two new 2-point functions, the one-loop correlation between matter displacements and the squared relative velocity, and the correlation function between the shear field $s_{ij}$ and the relative velocity. We describe these in turn.

The second order solution to the total-matter displacement (correct up to first order in the relative component) is given by 
\begin{equation}
    \bPsi^{(2)}_i(k) = \frac{1}{2} \frac{3}{7} \frac{i \, k_i}{k^2} \int \frac{d^3 p}{(2\pi)^3} \Big[ 1 - \Big( \frac{(k-p) \cdot p}{|k-p| |p|} \Big)^2    \Big] \, \delta_{m,0}(p) \, \delta_{m,0}(k-p),
\end{equation}
and more simply the ``normalized'' relative velocity at first order is given by 
\begin{equation}
   \vb{r}_{-,i}(k) = \frac{-i \, k_i}{k^2} \big( \nabla \cdot \vb{r}_{-} \big) (k).
\end{equation}
From this we can calculate the two-point function
\begin{align}
    \langle \bPsi^{(2)}(q)\ \vb{r}_{-}^2(0) \rangle = \frac{3}{7} \int \frac{d^3 k}{(2 \pi)^3} \, &e^{i k \cdot q} \Big( \frac{-i \, k_i}{k^2} \Big) \frac{d^3 p}{(2 \pi)^3} \nonumber \\
    &\Big[ 1 - \Big( \frac{(k-p) \cdot p}{|k-p| |p|} \Big)^2    \Big] \frac{p \cdot (k-p)}{p^2 (k-p)^2} P_{\delta_m \nabla \vb{r}_-}(p) P_{\delta_m \nabla \vb{r}_-}(k-p),
\end{align}
which can be simplified to give
\begin{equation}
    \langle \bPsi^{(2)}(q) v_{r,0}^2(0) \rangle = \hat{q} \int \frac{d^3 k}{(2 \pi)^3} \, e^{i k \cdot q} Q_{v^2}(k),
\end{equation}
where the kernel is defined as
\begin{equation}
    Q_{v^2}(k) \equiv \frac{3}{7}\int_0^\infty \mathrm{d} r\ P_{m-}(k r) \ \int_{-1}^1 \frac{\mathrm{d}x}{4 \pi^2} \frac{r\ (x-r)(1-x^2)}{(1+r^2-2 r x)^2}\ P_{m -}(k \sqrt{1+r^2- 2 r x}).
\end{equation}

Next, the shear-velocity correlation function $W^{s-}_{ijk}$ is given in Fourier space by
\begin{equation}
W^{s-}_{ijk}(q) = i \int \frac{d^3 k}{(2\pi)^2}\ e^{i k \cdot q} \Bigg( \frac{k_i k_j k_k}{k^4} - \frac{1}{3} \delta_{ij} \frac{k_k}{k^2} \Bigg) P_{m-}(k) \equiv \tilde{W}^{s-}_{ijk}(q) - \frac{1}{3} \delta_{ij} U^{m-}_k(q).
\end{equation}
where in the last equality we have split $W^{s-}_{ijk}$ into a totally-symmetric piece and a familiar piece proportional to $U^{m-}$. The former can be decomposed into scalar components
\begin{equation}
    \tilde{W}^{s-}_{ijk}(q) = \mathcal{A}(q)\ \hat{q}_i \hat{q}_j \hat{q}_k + \mathcal{B}(q)\ (\hat{q}_i \delta_{jk} + \hat{q}_j \delta_{ki} + \hat{q}_k \delta_{ij}),
\end{equation}
with the scalar components defined as spherical Bessel transformations:
\begin{equation}
    \mathcal{A}(q) = \int \frac{dk\ k}{2 \pi^2}\ j_3(kq)\ P_{m-}(k)
\end{equation}
\begin{equation}
    \mathcal{B}(q) = -\int \frac{dk\ k}{2 \pi^2}\ \frac{1}{5}\ \Big(j_1(kq) + j_3(kq)\Big)\ P_{m-}(k).
\end{equation}

\bibliographystyle{JHEP}
\bibliography{main}
\end{document}